\documentclass[letterpaper,twocolumn,10pt]{article}
\usepackage{usenix-2020-09}

\usepackage{tikz}
\usepackage{amsmath}
\usepackage{comment}

\usepackage{multirow}
\usepackage{pifont}
\usepackage{subcaption}
\usepackage{soul}
\usepackage{stfloats}

\usepackage{graphicx}
\usepackage{amssymb}
\usepackage[linesnumbered,ruled,vlined]{algorithm2e}
\SetKwInput{KwInput}{Input}                % Set the Input
\SetKwInput{KwOutput}{Output} % set the Output

\usepackage{color}

\usepackage{filecontents}
\usepackage{listings}
\usepackage{xcolor}

\usepackage[normalem]{ulem} %strikethrough

\usepackage{etoolbox}
\makeatletter
\patchcmd{\maketitle}
{\@maketitle}
{\@maketitle\vspace{-6em}}% change the value as needed

\newcommand{\pn}{{SuperScaler}}%\textsc
\newcommand{\plcode}{{sProgram}}
\newcommand{\coshard}{{co-shard}}

\newcommand{\ptensor}{{pTensor}}%\textsc
\newcommand{\textot}[1]{{#1}}

\newcommand*\circled[1]{\tikz[baseline=(char.base)]{
		\node[shape=circle,draw,inner sep=1.pt] (char) {#1};}}

\definecolor{forestgreen}{rgb}{0.13, 0.55, 0.13}

\newcommand{\TODO}[1]{{\color{red} {\bf #1}}}

\definecolor{dkgreen}{rgb}{0,0.6,0}

\newcommand{\eg}{{\it e.g.,\ }}

\newcommand{\ie}{{\it i.e.,\ }}

\newcommand{\para}[1]{{\bf \noindent #1 \hspace{2pt}}}

\definecolor{codegreen}{rgb}{0,0.6,0}
\definecolor{codegray}{rgb}{0.5,0.5,0.5}
\definecolor{codepurple}{rgb}{0.58,0,0.82}
\definecolor{backcolour}{rgb}{0.95,0.95,0.92}

\lstdefinestyle{mystyle}{
	backgroundcolor=\color{backcolour},   
	commentstyle=\color{codegreen},
	keywordstyle=\color{magenta},
	numberstyle=\tiny\color{codegray},
	stringstyle=\color{codepurple},
	basicstyle=\ttfamily\footnotesize,
	breakatwhitespace=false,         
	breaklines=true,                 
	captionpos=b,                    
	keepspaces=true,                 
	numbers=left,                    
	numbersep=5pt,                  
	showspaces=false,                
	showstringspaces=false,
	showtabs=false,                  
	tabsize=2
}

\date{}

\title{\vspace{-4cm} \pn{}: Supporting Flexible DNN Parallelization via a Unified Abstraction}

\author{
{\rm Zhiqi Lin$^{\dag,\ddag}$\thanks{This work was done when the authors were with Microsoft Research.} , Youshan Miao$^{\ddag}$, Guodong Liu$^{\S,\ddag}${\color{green}$^{*}$}, Xiaoxiang Shi$^{\P,\ddag}${\color{green}$^{*}$}, Quanlu Zhang$^{\ddag}$, Fan Yang$^{\ddag}$}, \\ 
{\rm Saeed Maleki$^{\ddag}$, Yi Zhu$^{\ddag}$, Xu Cao$^{\ddag}$, Cheng Li$^{\dag}$, Mao Yang$^{\ddag}$, Lintao Zhang$^{\ddag}$, Lidong Zhou$^{\ddag}$} \\ 
$^{\dag}$University of Science and Technology of China, $^{\ddag}$Microsoft Research, \\
$^{\S}$Institute of Computing Technology, Chinese Academy of Sciences, $^{\P}$Shanghai Jiao Tong University
}

\begin{document}
%-------------------------------------------------------------------------------

\maketitle

%-------------------------------------------------------------------------------
\begin{abstract}
%-------------------------------------------------------------------------------

\noindent With the growing model size, deep neural networks (DNN) are increasingly trained 
over massive GPU accelerators, which 
demands a proper parallelization plan that 
transforms a DNN model into fine-grained tasks and then schedules them to GPUs for execution. 
Due to the large search space, 
the contemporary 
parallelization plan generators often rely on empirical rules that couple transformation and scheduling, and fall short in exploring more flexible schedules that yield better memory usage and compute efficiency. This tension can be exacerbated by the emerging models with increasing complexity in their structure and model size.

\pn{} is a system that facilitates the design and generation of highly flexible parallelization plans.
It formulates the plan design and generation into three sequential phases explicitly: 
model transformation, space-time scheduling, and data dependency preserving.
Such a principled approach decouples multiple seemingly intertwined factors and enables the composition of highly flexible parallelization plans. As a result, \pn{} can not only generate empirical parallelization plans, but also construct new plans that achieve up to 3.5$\times$ speedup compared to state-of-the-art solutions like DeepSpeed, Megatron and Alpa, for emerging DNN models like Swin-Transformer and AlphaFold2, as well as well-optimized models like GPT-3.

\end{abstract}

\section{Introduction}
\label{sec:intro}

% Paragraph: models are getting bigger and we need parallelism
The size of deep neural network (DNN) has grown significantly~\cite{gpt-3, swin-v2, palm} over the past few years and the trend
clearly shows little sign of any slow down. Accelerators such as GPUs are crucial for training such models thanks to the compute power they offer. However, single GPU's memory has not scaled as fast as the model sizes.
Large DNN models like GPT-3~\cite{gpt-3} cannot fit into a single accelerator (\eg GPU) due to 
limited available memory. 
Therefore, utilizing parallel GPUs to distribute a model's weights has become the main method to enable training of such large models. However, efficiency on a distributed GPU cluster is a major DNN system research problem. 

% Paragraph: what is a DFG
Frameworks such as PyTorch~\cite{PyTorch} and TensorFlow~\cite{tensorflow} have  simplified expressing the architecture of a DNN in terms of basic operators such as matrix multiplication. Composing
these operators creates a data flow graph (DFG) which is a directed acyclic graph (DAG) and each node is a basic operator and every edge corresponds to the data dependency between the source and the destination. A DNN framework takes the DFG as an input and computes the operators following DFG dependencies. For a given model, this DFG is executed numerous times each with a different input and the model weights are updated with every few iterations.

% Paragraph: Why does a DFG need parallelization and how it is done? 
The corresponding DFG for a large model
may have large operators and a large graph. Limited compute power and memory capacity of a single GPU dictates partitioning large operators into multiple smaller independent operators 
and then assigning them to different GPUs.
%This computation process can execute in a single GPU. Parallel deep learning training identifies and exploits the inherent parallelism in the DNN model to speed up the training by utilizing multiple GPUs. 
%This involves partitioning a DNN model graph into smaller parts, which will be mapped and scheduled on the GPUs in a way that respects data dependencies in the original data flow graph and thus ensuring correctness. 
We  call the end-to-end scheme of partitioning and scheduling the DFG on multiple GPUs a {\em parallelization plan}.

% paragraph 
Finding an optimal parallelization plan is the cornerstone of efficient DNN training. For example, one has to evaluate the trade-offs between a spatial scheduling that computes a DNN model on multiple GPUs concurrently to improve the degree of parallelism and a temporal scheduling that runs operators of the model in one GPU sequentially to save communication costs~\cite{flexflow, tofu}. 
%as well as the fine grained mapping between pipeline stages (\eg dependent operators) and GPUs for better resource utilization~\cite{gpipe, terapipe}.
Similarly, the granularity of operator aggregation and mapping them to pipeline stages on different GPUs~\cite{gpipe, terapipe} is equally important to balance for optimal performance. 
At the same time, the plan should respect the original data dependency for correctness. In another word, the design of a parallelization plan requires a complex joint consideration of multiple intertwined factors: model partitioning, space-time scheduling (defining which GPU and when an operator executes), and data dependency preserving. To resolve this high complexity, existing systems support the parallel training of popular DNN models through predefined, well-studied parallelization plans composed by empirical parallelization rules, \eg tensor parallelism~\cite{megatron1, flexflow} or pipeline parallelism~\cite{dapple, gpipe}, or the combination of rules~\cite{megatron2, alpa, piper}. Such abstracted approaches show limited flexibility and do not work well on emerging DNN models like AlphaFold2 (\S\ref{sec:background}).

This paper presents \pn{}, a system that helps developers to design and generate highly flexible parallelization plans for deep learning training. Departed from empirical solutions, \pn{} takes a principled approach that formulates the design of parallelization plan as three sequential phases explicitly: model partitioning, space-time scheduling, and data dependency materialization. 

% \cheng{I have a hard time to learn insights from the following technical contributions. I suggest that we need to highlight two messages: first, flexibility is achieved by (1) the decoupled abstraction, which allows users to consider transformation and scheduling separately, and (2) the simple, fine-grained primitives for the two steps to specify something that existing solutions do not support. For instance, with new scheduling primitives, we are able to specify ordering constraints between operators which may come from different parallelisms. }

% \cheng{The second message is that flexibility is a double-edge sword. The specified scheduling can be wrong as it does not comply with the specified transformation or violates data dependencies. Therefore, we propose an automatic reasoning method, which tracks data dependencies and something else to find conflicts. This alleviate the programming burden on the users' shoulder.}

In the model partitioning phase, \pn{} provides \texttt{op-trans}, a primitive that allows developer to express model partitioning as the transformation of each operator in the data flow graph representing the model. The developer can provide multiple legal transformations for one operator, and \pn{} can compose them into graph-level transformation. % while tracking the logical data dependency before and after the transformation. Such graph composition provides a highly flexible way of DNN model partitioning, which in turn leads to highly flexible scheduling options. 
Given a transformed graph (partitioned model), \pn{} moves to the scheduling phase. 
% It provides \texttt{op-sched}, another set of primitives offered to developers to express various space-time scheduling schemes by mapping a portion of partitioned model to a certain GPU spatially. The \texttt{op-sched} contains a \texttt{happen-before} primitive to enforce the temporal execution order between operators without data dependency. 
It provides \texttt{op-assign} and \texttt{op-order} primitives for developers to express various space-time scheduling schemes, with \texttt{op-assign} mapping a portion of partitioned model to a certain GPU spatially, and \texttt{op-order} expressing a \texttt{happen-before} constraint to enforce the temporal execution order between operators without explicit data dependency. These two phases allow developers to consider transformation and scheduling, \emph{separately} which enables the expression of highly flexible parallelization plans that existing solutions do not support (\S\ref{subsec:example-para}). 
%For instance, the new \texttt{op-order} primitive can specify ordering constraints between operators coming from different pipeline stages (\S\ref{subsec:example-para}). %enables flexible temporal scheduling required by various pipeline parallelism schemes. Given a scheduling scheme, \pn{} also performs deadlock detection through the analysis of data dependency tracked during graph transformation to reduce scheduling mistakes. 

The flexibility enabled by the separation between model partitioning and scheduling may increase the burden of developers as the transformation and scheduling process could be error-prone. The transformation may require sophisticated changes in data dependencies to preserve the correct mapping between new and original graphs. For example, one may accidentally specify temporal scheduling order that violates data dependencies and leads to deadlocks. To address this problem, \pn{} introduces vTensor to track the logical data dependencies before and after each operator transformation, and maintains the dependencies between transformed operators through the original data flow graph (\S\ref{subsec:model-trans}). After scheduling decision is made, \pn{} performs deadlock detection through the analysis of the tracked data dependency, alerts the developers of potential violation so that they can refine the design accordingly. The process repeats iteratively until no violation is detected. This facilitates the reasoning of various parallelization plans during the design process.  

%Once model partitioning and space-time scheduling are expressed through \texttt{op-trans} and \texttt{op-sched}, 
In the final phase, \pn{} will automatically materialize the logical data dependency tracked during graph transformation and scheduling. 
%perform data dependency adaptation to ensure the new transformed and scheduled graph preserves the data dependency of the original DNN model. 
The dependency materialization automatically inserts a collective communication primitive such as all-reduce for an operator that is split and scheduled across GPUs. These communication primitives can have unconventional semantics if two dependent operations are assigned to two different number of GPUs (\S\ref{sec:comm-opt}).
%It can also introduce a new operator inserted between an upstream operator and downstream operator to remap the mismatched tensors if the two operators are partitioned into different number. 
%Such materialization is only possible after the transformation and scheduling are done. 
The automatic data dependency materialization and communication operation insertion further relieves developers from the tedious and error-prone process of exploring parallelization plans. 
%Due to the flexibility, the materialization process may exploit unconventional communication patterns to optimize the performance (\S\ref{sec:comm-opt}).

With the above design, \pn{} decouples multiple seemingly intertwined factors and enables developers to design different parallelization plans without worrying about the underlying system implementation details.

\pn{} is implemented based on PyTorch~\cite{PyTorch}. Developers use \pn{}'s primitives, \texttt{op-trans}, \texttt{op-assign} and \texttt{op-order}, to write \plcode{}, a program describing how a given DNN model, represented by a DFG, is transformed and scheduled. \pn{} will then compile the \plcode{} into 
an execution flow graph served as an intermediate representation of the parallelization plan. \pn{} 
% analyzes the data dependency of this graph to detect possible deadlocks. 
analyzes this graph for automatic data dependency materialization and deadlocks detection.
Finally, the resulting \pn{} graph is compiled back into PyTorch codes for execution via PyTorch engine. 

The flexibility of \pn{} offers easy exploration of new parallelization plans such as \coshard{} and interlaced pipeline (\S\ref{sec:background}) besides existing empirical plans. This is only possible by \pn{}'s flexible space-time scheduling and materialization of data dependency leveraging unconventional communication patterns.
The resulting parallelization plans are shown to achieve 3.5$\times$ speedup compared to state-of-the-art parallel training systems, including DeepSpeed~\cite{deepspeed}, Megatron~\cite{megatron2} and Alpa~\cite{alpa}, for emerging DNN models in computer vision (Swin-Transformer~\cite{swin-v2}), language translation (mBART~\cite{mbart}), and biology analysis (AlphaFold2~\cite{alphafold}). Surprisingly, with the new parallelization plans \pn{} can even achieve 1.5$\times$ speedup for well-optimized models like GPT-3 under certain realistic configurations. We plan to release \pn{} to the open-source community.
\section{Background and Motivation}
\label{sec:background}

\begin{figure}[t]
    \centering
    \includegraphics[width=1.0\linewidth]{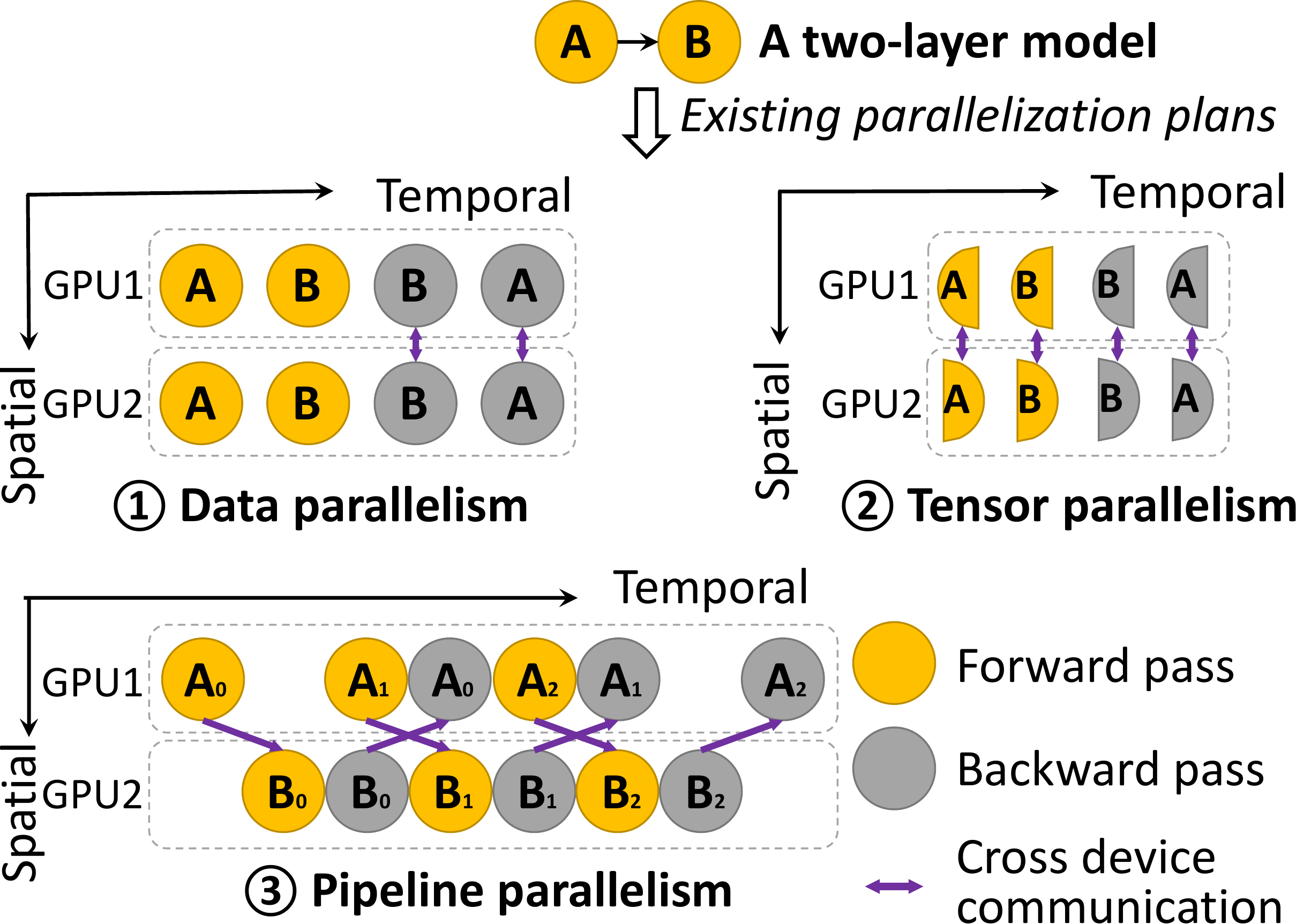}
    % \vskip -1ex
    \caption{Empirical parallelization plans. A$_i$ and B$_i$ denote the computation on $i$-th micro-batch.}
    % \vskip -2ex
    \label{fig:parallelplans}
\end{figure}

\para{Empirical parallelization plans.}
%As deep learning models are increasingly large, distributed execution of model training and inference becomes a must. There has been some empirically designed parallelization plans as shown in \autoref{fig:parallelplans}. 
Today, parallel deep learning training mostly relies on parallelization plans composed by empirical rules. Figure~\ref{fig:parallelplans} highlights several empirical parallelization plans. As most DNN models have multiple layers, \emph{data parallelism} partitions all the layers to the same number of pieces along the batch dimension of the input tensor, the partitions are spatially placed on disjoint devices. Cross-partition communication only happens during the backward pass, a computation executes in the reverse direction of the data dependency depicted in the model's data flow graph, using the ``all-reduce'' primitive. Data parallelism assumes the whole model can fit in a single device, which is often not feasible for large models. Thus, \emph{tensor parallelism} partitions a model (usually along an axis other than the batched dimension) evenly and places on disjoint devices. In tensor parallelism, cross-partition communication is complicated, it happens during both forward and backward passes, and requires different types of communication primitives (\eg all-reduce, reduce-scatter). This parallelization plan often introduces a large amount of communication. \emph{Pipeline parallelism} can greatly reduce communication costs in some popular models (\eg BERT~\cite{devlin2018bert}). It groups layers of a model into several stages, each is placed on a dedicated device(s). The execution of each stage is temporally scheduled by pipelining the execution of multiple micro-batches, \ie splitting a mini-batch into multiple micro-batches, as shown in Figure~\ref{fig:parallelplans}. The communication across devices is peer-to-peer send/receive. Some improve the device utilization, some propose to improve the naive pipeline parallelism with more sophisticated temporal ordering across micro-batches (\eg 1F1B~\cite{dapple}). % its own flaws, \eg the idle cycles in pipeline, imbalanced memory consumption. To move a step further for better performance, 
Moreover, the above empirical parallelization plans can be combined in a sequential and nested way, \ie pipeline parallelism with data or tensor parallelism applied within each pipeline stage~\cite{megatron1, gspmd, alpa}, to further improve performance.
% to provide further performance improvement.

Although effective for popular DNN models, all these empirical parallelization plans couples model partitioning, space-time scheduling, and the corresponding changes in data dependency together. For example, both data and tensor parallelisms carefully align the communication primitives with the way the model is partitioned and assigned to disjoint devices. Pipeline parallelism requires careful temporal ordering to minimize device idle time, assuming stages are placed disjointly. This practice limits the flexibility of parallelization plans, as elaborated next.

\begin{figure*}[t]
    \centering
    \includegraphics[width=0.99\linewidth]{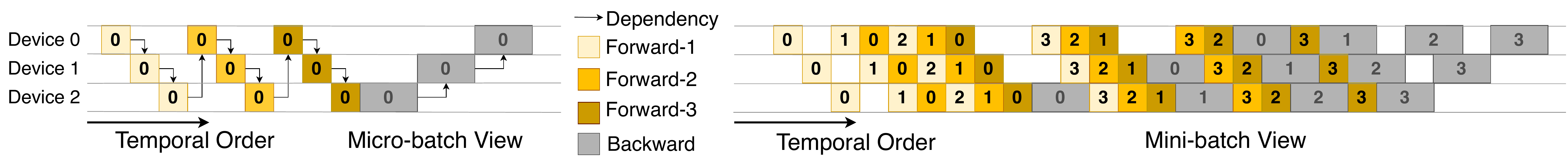}
    \caption{A specialized parallelization plan for AlphaFold.}
    \vskip -2ex
    \label{fig:alphafold}
\end{figure*}

%\para{Specialized parallelization plan boosts efficiency.}
\paragraph{Emerging DNN models require new parallelization plans.}

We observe that new DNN models have unique characteristics that existing parallelization plans cannot support effectively.
For example, AlphaFold2~\cite{alphafold}, the state-of-the-art model for molecular structure analysis, has three forward passes and one backward pass. As shown on the left of Figure~\ref{fig:alphafold}, the output of each forward pass is the input of next pass, all the way to the backward pass. To effectively support AlphaFold2, the model needs to be trained in pipeline parallelism, but with a new temporal ordering specified for the micro-batches (shown on the right of Figure~\ref{fig:alphafold}). This new pipeline parallelism pattern (we call it 3F1B) cannot be expressed by any of the existing pipeline schemes. And to define a new rule for this pattern clearly lacks flexibility. This motivates us to design a new scheduling primitive for order preservation. 

\begin{figure}[t]
    \centering
    \includegraphics[width=1.0\linewidth]{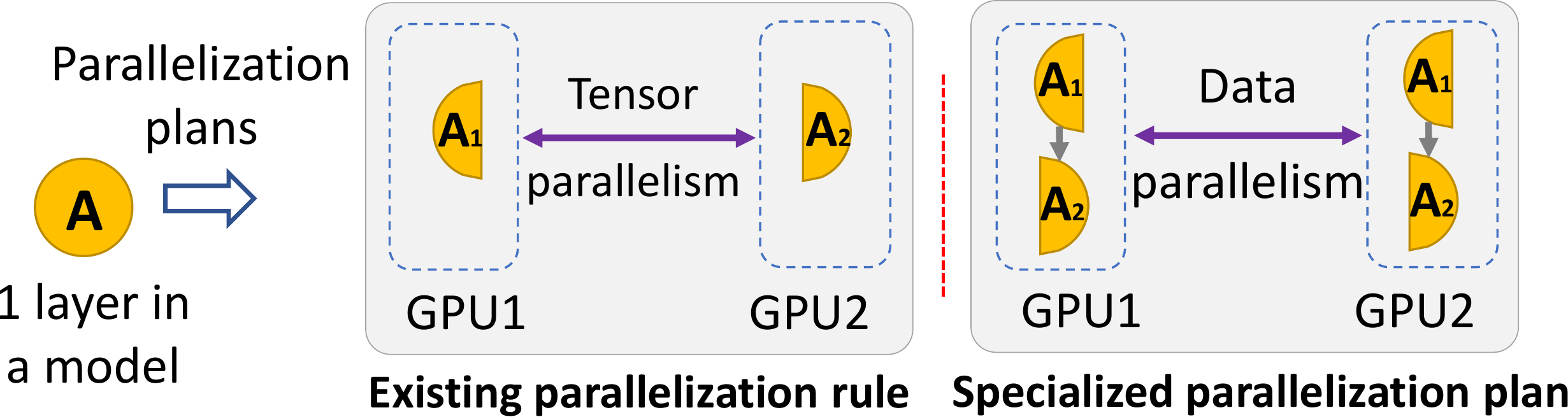}
    \caption{\coshard{}: a new parallelization plan.}
    % \vskip -1.5ex
    \label{fig:coshard-lite}
\end{figure}

Another example is shown in Figure~\ref{fig:coshard-lite}. This new parallelization plan, \coshard{} breaks the assumption that model partitions are required to be mapped to disjoint devices. The diagram on the right of Figure~\ref{fig:coshard-lite} shows that partitioned model can be placed to the same GPU and executed sequentially. This makes it possible to apply more communication efficient data parallelism across GPUs. In contrast, the traditional tensor parallelism could introduce higher communication cost. Co-shard is shown effective for Swin-Transformer\cite{swin-v2}, a new transformer for vision tasks. Co-shard partitions the model along its multi-head dimension~\cite{attention}, co-locates the partitioned model in one GPU, leveraging recompute~\cite{recompute} to reduce the peak memory usage, and achieves up to 3.5$\times$ speedup (\S\ref{sec:eval}).

\S\ref{subsec:example-para} discusses some more examples. All these highlight the limitation of empirical parallelization plans and necessitate a system to facilitate the design of highly flexible parallelization plans.

\section{Design}
\label{design}

% In this section, we introduce the design of \pn{} that effectively formalizes DNN model parallelization into the three sequential phases: model transformation, space-time scheduling and data materialization.

\begin{figure}
% \begin{lstlisting}[numbers=none]
% DNN Model (XFG) =>

% 1) model transformation 
% => XFG exploit parallel
% 2) space-time scheduling 
% => XFG with validated scheduling
% 3) data materialization
% => XFG with data tensor/communication connecting operators

% => ready to compile for execution
% \end{lstlisting}
	\centering
	\includegraphics[width=.8\linewidth]{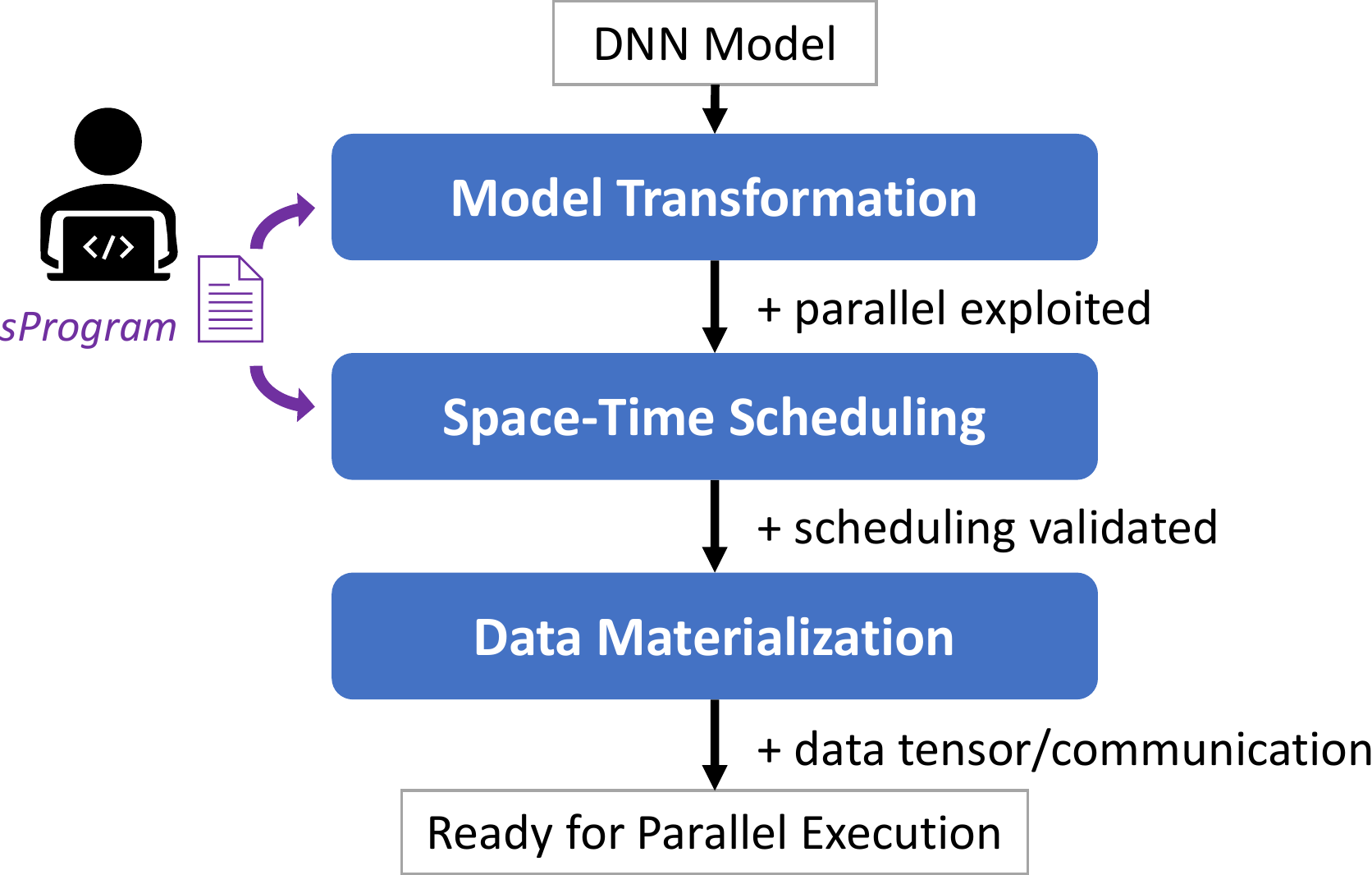}
        % \vskip -1ex
	\caption{\pn{} system workflow.}
        % \vskip -2ex
	\label{fig:sys-overview2}
\end{figure}

% \zhiqi{Follow the above motivation, We design \pn{} by allowing system expert to focus on the expression of transformation and scheduling on computation graph, while delay the data dependency materialization on system side until the finish of them. Figure~\ref{fig:sys-overview2} illustrates the overall workflow of \pn{}. To achieve this goal, system firstly takes a DNN model described in traditional data flow graph (DFG), and convert it to execution flow graph (XFG), which only tracks data dependency but allows not to materialize it. Then, the XFG is used as input for sProgram, where system experts specify operator transformation and scheduling on it. After the specification, system automatically materializes the data dependency by inserting operators to guarantee tensor consistency between operators. Finally, the XFG are delegated to devices for execution.}

%Following the motivation, 
To support more flexible parallelization plans, \pn{} allows developers to focus on model partitioning and space-time scheduling, while delegating the sophisticated, error-prone process of data dependency materialization to \pn{}.
Figure~\ref{fig:sys-overview2} summarizes the overall workflow of \pn{}.
The input of \pn{} is a DNN model computation graph.
Besides DNN model, developers also provide an \plcode{} that express a parallelization plan with primitives \texttt{op-trans}, \texttt{op-assign} and \texttt{op-order}.
% trans
\pn{} first exploits inherent parallel of DNN model by applying  \texttt{op-trans} to partition operators into multiple functional equivalent operators.
\pn{} also tracks the data relations during model transformation (\S \ref{subsec:model-trans}).
% data dependency information.
% for next space-time scheduling phase and dependency materialization phase.
% sched
Then, \pn{} performs space-time scheduling with primitive \texttt{op-assign} assigning each operator an execution device and \texttt{op-order} that enforces execution orders between operators.
\pn{} also builds data dependency from tracked data relation to 
validate scheduling and avoid possible deadlock (\S \ref{subsec:model-sched}).
% materialize
Finally, \pn{} materializes the tracked data dependency into communications that connect mismatched data partitioning and cross device operators, and generates parallel execution (\S \ref{subsec:data-material}).

\subsection{Operator Transformation}
\label{subsec:model-trans}

% \begin{figure}[th]
% 	\centering
% 	\includegraphics[width=\linewidth]{figures/xfg-crop.pdf}
% 	\caption{Traditional DFG vs. XFG.}
% 	\label{fig:xfg-design}
% \end{figure}

DNN models, defined as operators performing computation over high dimensional tensor data, can be partitioned into finer-grain tasks to exploit parallelism.
\pn{} performs such operation over each operator with \texttt{op-trans}. 
% that substitutes a single operator into a set of functional equivalent operators.
Following a user-defined transformation algorithm \texttt{algo}, an \texttt{op-trans(op, algo)} partitions an operator \texttt{op} along with its input and output data tensors into a set of functional equivalent operators and tensors.
% Performing \texttt{op-trans} on an operator affects tensor number and shape, breaks the data connection with its consecutive operators, leading to insert new operations as a fix.
% In Figure~\ref{fig:xfg-design}, the DFG operator $A$ and its output get partitioned into $A_{1}$, $A_{2}$ and $Tensor_{1}$, $Tensor_{1}$, respectively.
% For the mismatched producer with consumer, an concat operation is inserted to merge $Tensor_{1}$, $Tensor_{1}$ and serve its downstream operator $B$ as input.

\begin{figure}[t!]
	\centering
	\includegraphics[width=.8\linewidth]{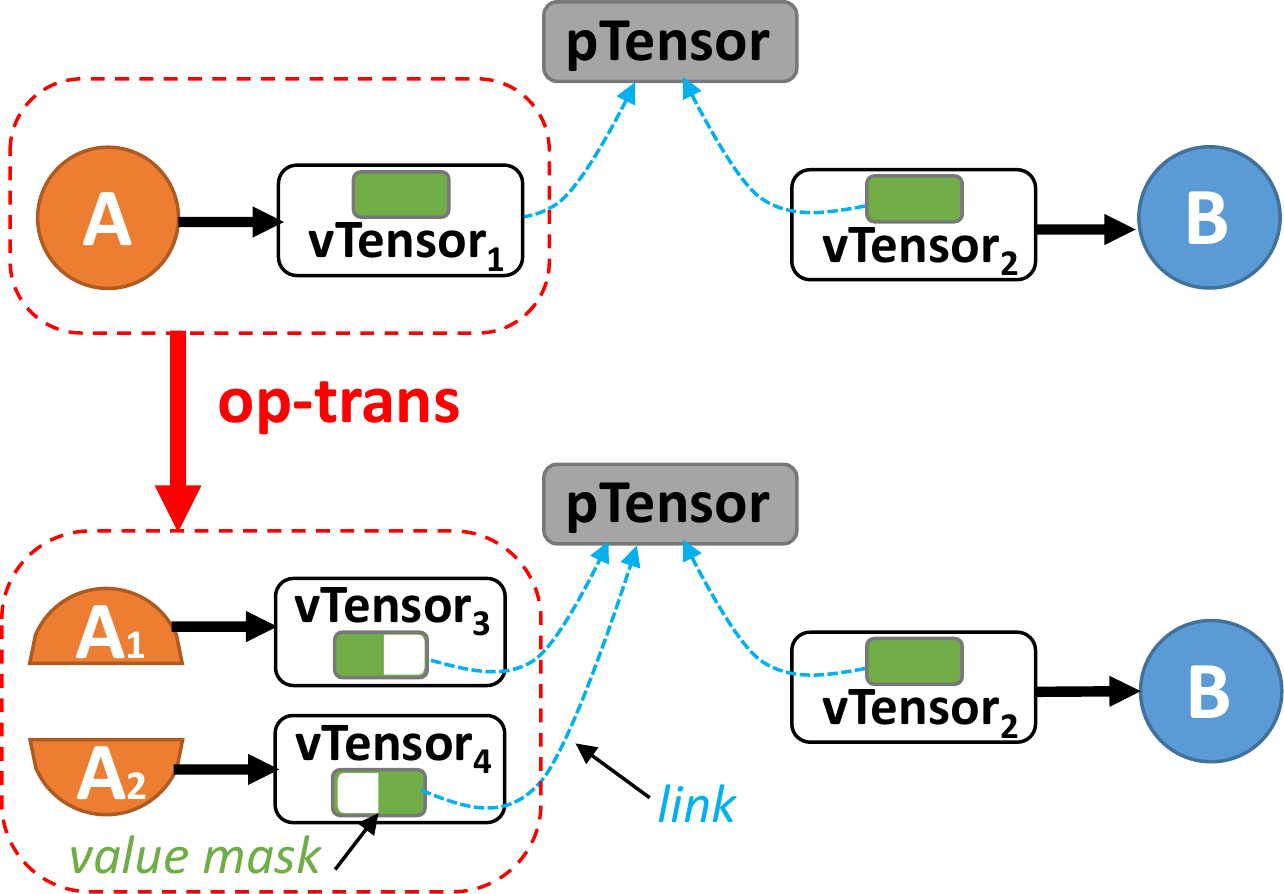}
	\caption{Perform \texttt{op-trans} on \pn{} graph.}
	\label{fig:xfg-design}
\end{figure}

\begin{figure}
    \centering
    \includegraphics[width=.85\linewidth]{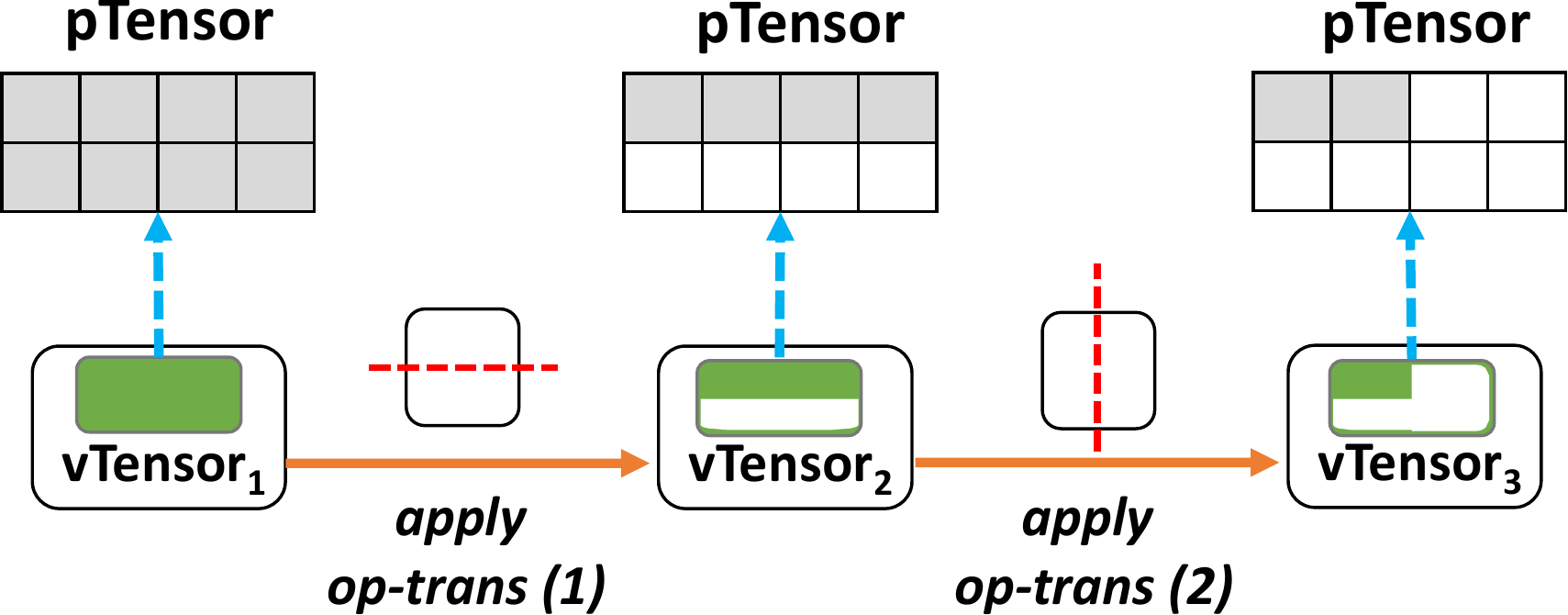}
    \caption{vTensors track data dependency during a series of \texttt{op-trans}.}
    % \vskip -3ex
    \label{fig:vtensor-remap}
\end{figure}

\para{vTensor.} 
\pn{} introduces vTensor to track the changing data dependency during operator transformation.
%For \texttt{op-trans} operation, we design vTensor as operators' input and output in \pn{} computation graph. 
A vTensor ``links'' to a \ptensor{}, which is a logically persistent tensor defined in the original DNN model (\eg Figure~\ref{fig:xfg-design}).
Besides the link, a vTensor also maintains a ``mask'', representing which portion of the \ptensor{} the operator accesses (\eg Figure~\ref{fig:vtensor-remap}).
Each operator has their own dedicated input and output vTensors, even multiple operators access the same \ptensor{}.
As shown on the top of Figure~\ref{fig:xfg-design}, operator \texttt{A}'s output data is operator \texttt{B}'s input. The two operators linked to the same \ptensor{} through their own vTensors, respectively.
Leveraging vTensor, a transformation algorithm in \texttt{op-trans} is defined as a graph substitution, which describing: 1) each new operator's computation, \eg MatMul, Add, and 2) how to partition original input and output vTensors to get new operators' input and output vTensors.
When applying an \texttt{op-trans}, \pn{} will only partition vTensors and leave \ptensor{}s unchanged.
And \texttt{op-trans} over an operator won't affect other operators' vTensors, as different operators have dedicated vTensors.
As shown in Figure~\ref{fig:xfg-design}, applying \texttt{op-trans} on operator \texttt{A} only splits itself and its output vTensor, leaving operator \texttt{B} unchanged.
Such separation allows developers to flexibly perform \texttt{op-trans} on different operators in \plcode{}. Moreover, developers do not need to align tensors between adjacent operators during transformation, \eg aligning vTensor$_3$ and vTensor$_4$ with vTensor$_2$, leaving such a tedious and error-prone process to the phase of data dependency materialization. 

\para{Data dependency tracking through vTensor.}
With vTensor, the link to pTensor, and the mask in a vTensor, \pn{} can track data dependency during operator transformations.
%Logical data-dependencies between operators is important for \pn{} to support scheduling and data materialization. \pn{} tracks data relation during \texttt{op-trans} and keeps in vTensors,
% in XFG that always keeping their data relation to cTensor, 
%which can later be used to rebuild data-dependency between operators.
When \texttt{op-trans} partitions a vTensor into multiple vTensors, each new vTensor links to the same \ptensor{} as the original vTensor but with a different mask.
%It makes the after \texttt{op-trans} vTensors still tracks their data access relation with \ptensor{}.
%If applied with another \texttt{op-trans} that further partitions these vTensors, such vTensor-\ptensor{} relation are still preserved.
%
% A vTensor in a op-trans tracks a step of tensor mapping from before to after the transformation.
% When an operator applied with multiple op-trans, there are multiple steps of mapping.
% Following the chain of mapping, we can accumulate the mapping effect to connect the final tensor with cTensor.
% In other words, despite applying any op-trans, we can always map a vTensor to a cTensor.
Figure~\ref{fig:vtensor-remap} shows how \pn{} tracks data dependency after two \texttt{op-trans}. %before \texttt{op-trans} partitioning any tensors, \textot{vTensor$_{1}$} maps to the entire \ptensor{}.
In the process, \texttt{op-trans(1)} partitions the tensor horizontally, the resulting \textot{vTensor$_{2}$} maintains a mask atop to show it keeps the top half of \textot{vTensor$_{1}$} to the pTensor.
%Then we can infer \textot{vTensor$_{2}$}'s map to \ptensor{} .
\texttt{op-trans(2)} further partitions \textot{vTensor$_{2}$} vertically, turning it into \textot{vTensor$_{3}$}, whose mask indicates that the left half of \textot{vTensor$_{2}$}, which is the top-left part of \ptensor{}.
%
%As \ptensor{} data structure in \pn{} graph immutable during \texttt{op-trans}, and vTensors always tracking relation with \ptensor{}, we can check data relation between two vTensors by comparing their value masks for the same \ptensor{}. 
For two vTensors linked to the same pTensor, \pn{} can easily detect whether they have data dependency by intersecting their masks. Such logical dependency will be used for space-time scheduling and dependency materialization.

\subsection{Space-Time Scheduling}
\label{subsec:model-sched}

% \zhiqi{suggest derive op-sched to op-assign and op-order. Otherwise the interface is same and hard to understand meaning.}

%Space-time scheduling defines the parallel execution scheme for an \pn{} graph, including operators' execution device and order. In an \plcode{}, space-time scheduling is expressed with \pn{}'s \texttt{op-assign} and \texttt{op-order} primitives. For flexibility consideration, similar to \texttt{op-trans} that performs transformations at operator granularity operations, \texttt{op-assign} and \texttt{op-order} also performs scheduling operations on operators.
%
\pn{} introduces \texttt{op-assign(op,device)} and \texttt{op-order(op1,op2)} to enable flexible space-time scheduling.
For example, \texttt{op-assign(op1, GPU0)} assigns device \texttt{GPU0} to execute operator \texttt{op1}. \pn{} records such assignment by annotating the data flow graph, which will be enforced during execution. After the assignment, the corresponding input and output tensors of the assigned operators naturally co-locate on the same device.
% \zhiqi{This design works well with memory optimizations like swap, which can be easily achieved by inserting an identity operator and assigning it to CPU.}
%
%The \texttt{op-order} primitive performs temporal scheduling by defining a happen-before relation between two operators. For example, 
\texttt{op-order(op1, op2)} adds a happen-before edge in \pn{} graph between the two operator nodes, and will perform \texttt{op1} computation before \texttt{op2} during execution.

\begin{figure}
    \centering
    \includegraphics[width=.95\linewidth]{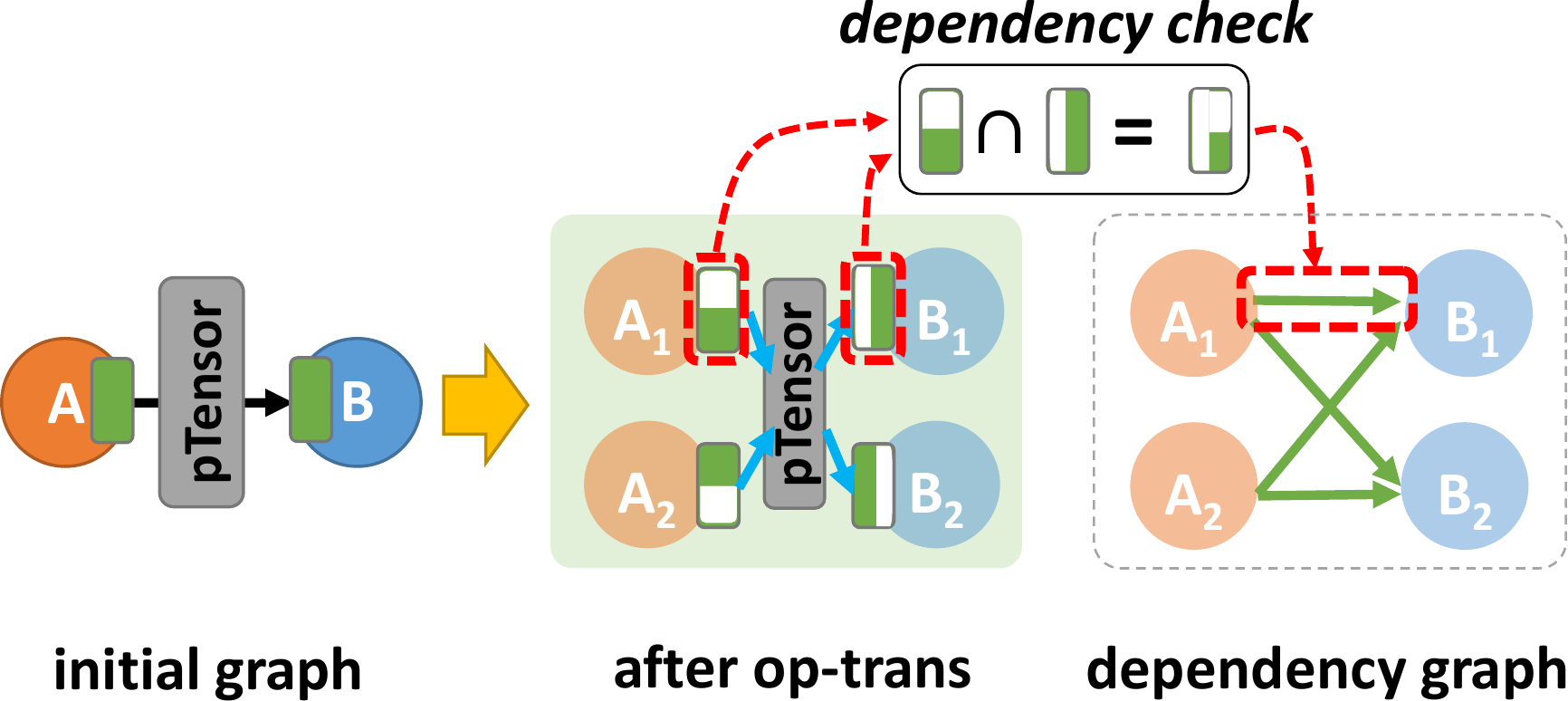}
    \caption{Checking data dependency between vTensors.}
    % \vskip -2ex
    \label{fig:depend-check}
\end{figure}

\para{Scheduling validation and completion.}
The freedom of arbitrary order specifying makes it possible  that some \texttt{op-order} calls violate previous \texttt{op-order}s or data dependency.
%It is possible that certain temporal scheduling \texttt{op-order} accidentally violates existing \texttt{op-order} or data-dependency.
To avoid potential deadlock and keep scheduling plans feasible, %we design a scheduling validation mechanism in \pn{}.
\pn{} performs scheduling validation as follows.
First, %scheduling validation extracts all the operator data-dependencies.
for each pair of producer and consumer operators in the initial graph,  \pn{} performs an interaction over their vTensor masks. Non-empty intersections indicate the existence of data dependency, as shown in Figure~\ref{fig:depend-check}. This way \pn{} can identify all data dependencies after op transformation and scheduling. 
%With dependency check over each ``producer consumer'' operator pair, we can find all data dependencies in \pn{} graph. 
% by: 1) computing the new tensor's mask over original tensor with the vTensor map; 2) comparing a consumer mask with a producer mask for intersection. And there is data-dependency if intersection exists.
With the identified data dependencies and the ``happen-before'' relations as edges, \pn{} will build a full dependency graph. And the execution scheduling is feasible only if it is an acyclic graph, which can be checked with graph circle detection algorithms~\cite{DFS}.
% alternative dependency
In certain cases, such as replicated producers, the consumer may depend on any one of the producers. \pn{} will enumerate these possibilities and consider the scheduling feasible if at least one acyclic graph exists.
% following one of possibilities leads to an acyclic graph.
% auto complete (to imple.?)
%
In some cases, the operator execution order on one device is unspecified, introducing ambiguity. %if the space-time scheduling does not determine sequential operator order on each device, it leaves different possible scheduling to execution. 
To avoid potential deadlock due to ambiguous execution, \pn{} specifies a feasible order for these operators by %, when finished applying all scheduling in \plcode{}.
applying a topological sort~\cite{toposort} over the full dependency graph and returning the global sequential order. %, making execution order on each device determined.
% \zhiqi{suggest the above part move in scheduling space. We can make a clean separation that scheduling space is about problem space, scheduling validation is about checking whether user specified constraints satisfy the problem space.}

%The space-time scheduling primitives allows developers flexibly expressing scheduling scheme and the scheduling validation enables \pn{} to make sure it is feasible to generate parallel execution.

% diff op-op data-dependency lead to different schedule possibility (illustrate)
%             scheduling limited by exsiting op-op order constraint 
%                 data-dependencies and existing "happen-before"

%     extract op-op data-dependencies
%         merge access-mapping (from multiple-times op-trans)
%         producer-consumer overlapping check
%     checking scheduling feasibility with, \eg topo-sort or loop detection //move to imple.?
%         false alarm at multi-alternative connections?

% key: with data-dependencies, it can check possible scheduling for current model transformation

\subsection{Dependency Materialization}
\label{subsec:data-material}

%Model transformation and scheduling determine parallelization plan.
%% But the XFG cannot execute directly as operators are still isolated, without any dataflow to connect the operators.
After transformation and scheduling, operators in a \pn{} graph may have some upstream output vTensor mismatched its downstream input vTensor, or located in different devices.
% \zhiqi{the key is to point out tensor shape / value mismatch.}
Data dependency materialization fixes these problems with the following steps. % and materialized dataflow and communications.

\begin{figure}
    \centering
    \includegraphics[width=.7\linewidth]{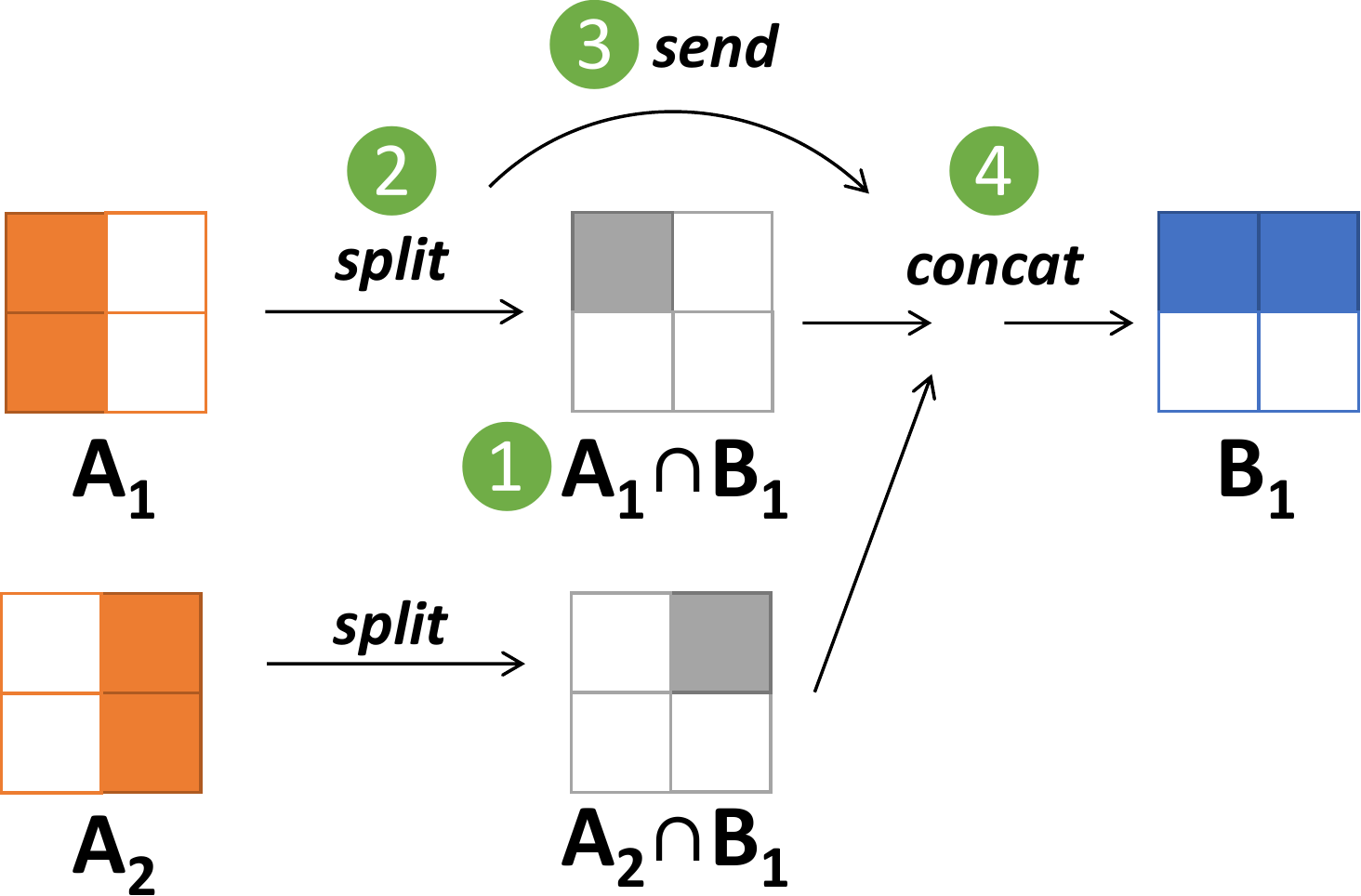}
    \caption{Data dependency materialization for producer vTensors $A_{1}$, $A_{2}$ and consumer vTensor $B_{1}$.}
    \label{fig:tensor-mater}
\end{figure}

First, the non-empty overlapped portion of producer and consumer vTensors are identified by intersecting the masks in both vTensors. Second, for the produced vTensor, a split operator is inserted to extract the overlapped portion of the two vTensors. A pair of send-recv operators will be inserted if the two vTensors locate in different devices. Finally, a concat or reduce operator on the consumer side is inserted to construct an input vTensor with the desired shape from multiple producers.
%Similar to scheduling validation, dependency materialization also leverages the data dependencies in \pn{} graph.
% 1)
%First, system performs a data-dependency check. Instead of only evaluating whether a pair of operators data-dependent or not, dependency materialization will also compute the ``overlapped'' region of the consumer tensor and producer tensor, to extract the dependent portion of the data. This is achieved with intersection operation over consumer and producer's vTensor masks.
%
% Then, with the overlapping information, adaption can 
% connect producer operators with consumer operators by generating appropriate operations
% such as slice and concat for dimensional division, reducing operations for partial value, and send-recv communication if cross-device.
% Specifically, according to above overlapped region check, we know which producers contains the data that a consumer operator needs as its input.
% 2)
%Second, we can add split operators to extract overlapped data region from producer operator's output tensors. 
% 3)
%Then, it inserts a pair of send-recv operations if consumers locates on a different device than producer. 
% 4)
%Finally, we add a concat or reduce operations on consumer side to construct its input tensor from pieces from different producers.
%
Figure~\ref{fig:tensor-mater} illustrates an example of data dependency materialization. Operators $A_{1}$ and $A_{2}$ produce the left and right half of the tensor and operator $B_{1}$ consumes the top half.
The overlapped regions $A_{1} \cap B_{1}$ and $A_{2} \cap B_{1}$ are the top-left and top-right part, respectively~\circled{1}.%, which from producer and will used by consumer operator \circled{1}.
It then splits both $A_{1}$ and $A_{2}$ to extract the overlapped region \circled{2}.
For cross-device tensors, communication operation will be added \circled{3}.
Finally, on the receiver side, it concats the collected pieces as $B_{1}$ \circled{4}.
% use a greedy algorithm to 
%Above generated operation will also be inserted into \pn{} graph as nodes for later code generation.
The changes will be recorded in the \pn{} graph for later code generation. During materialization, there exist optimization opportunities for communications, which we elaborate in \S\ref{sec:comm-opt}.
%Such approach can materialize dependencies for any cases, and the efficiency will be optimized in \S\ref{sec:comm-opt}.

% extract consumer-producer relation
%     (extract op-op data-dependencies)
%     for dependent producer-consumer pairs, perform data region overlapping check
%         choosing from multi-alternative sources?
    
% adaption generation
%     straw-man approach (illustrate): 
%         slice, (send-recv if cross device), reduce/concat        
%     we will optimization (\S \ref{sec:imple})

% key: already make sure model transformation and scheduling are legal, it only needs to perform tensor/comm. materialization according to data-dependency

\subsection{Exploring More Parallelization Plans}
\label{subsec:example-para}

With the above design, \pn{} can support existing popular parallelization plans as well as new flexible parallelization plans for emerging models. Note that we skip the new pipeline parallelism for AlphaFold2 and \coshard{} as they have been discussed in \S\ref{sec:background}. 

% classic parallelism
%     data-parallel
%     pipeline-parallel
%     recompute

% new parallel plans
%     \coshard{}
%     mbart

\subsubsection{Data Parallelism}
% \paragraph{Example \plcode{}}
% In \pn{}, with these concise and flexible interface and primitive, a user can easily express model-execution mapping including classic parallelisms. 
% Figure~\ref{fig:data-parallel}
Algorithm~\ref{alg:data-parallel-sprogram} shows an example \plcode{} for data parallelism. 
It takes a \pn{} graph and device environment as input.
% converted from the logical model DFG.
% An operator in DNN has certain roles, i.e., data-loader, forward computation and optimizer, which is preserved in XFG to help identify different operators.
% For each operator in the XFG, it performs op-trans with different algorithms according to its role. 
% Specifically, in data-parallelism \plcode{}, it uses `replicate' op-trans to duplicate optimizer operators and `split' op-trans on the rest operators to dividing along samples dimension.
Each forward computation operator will be partitioned along the ``batch'' dimension with \texttt{op-trans} (Line 3-5). The other optimizer operators will be replicated (Line 6-7).
Then the transformed operators will be assigned among devices (Line 8-9).
The operator type and dimension information used in \texttt{IsForward} and \texttt{GetBatchDim()} are captured from DFG and kept in \pn{} graph (\S\ref{sec:imple}).
Note that backward operators can be omitted in the specification, \pn{} will adapt them to their forward operators automatically through operator transformation (\S\ref{sec:imple})
% \pn{} leverages the named tensor technology~\cite{named-tensor} to provide dimension information such as `batch' in XFG. \zhiqi{this looks not much relavant about our core.}
% More details about transformation algorithm will be introduced in~\S\ref{subsec:prog2exe}.
% Then, with \texttt{op-assign} primitive, the new operators are assigned among all devices evenly to execute different sub-batch of samples concurrently.

% \zhiqi{Algorithm~\ref{alg:data-parallel-sprogram} shows an example of expressing data-parallelism. Each forward operators in the graph can be partitioned alongside the batch dimension (Line 3-5). Optimizers are transformed with replication (Line 6-7). The transformed operators are then assigned to different devices(Line 8-9). Note backward operators can be omitted in the specification due to our provided facilities of automatic transformation with their forward operators (\S\ref{sec:imple})}

% \zhiqi{I suggest to remove optimizers, it looks we are making pitfalls. We don't include it in other sProgram. If we remove it, we can change the data parallelism sProgram to a more general tensor parallelism sProgram, and compare it with Coshard to demonstrate the key message: one transformation has multiple scheduling choices, and is easy to express in our principle.}

\begin{algorithm}[t]
\DontPrintSemicolon
   \small
   \KwInput{\pn{}Graph g, Environment env}
   \KwOutput{transformed \pn{}Graph g}
   
   ndevs $\gets$ $\lvert$env.devices$\rvert$ \tcp*{get device number}

   \For{op $\in$ g.ops}
   {
     \uIf(\tcp*[f]{partition forward ops}){\upshape IsForward(op)}
     {
       dim $\gets$ GetBatchDim(op) \\
       new\_ops $\gets$ \textbf{op-trans}(op, SplitAlgo(dim, ndevs)) \\
     }
     \Else(\tcp*[f]{replicate optimizer ops})
     {
       new\_ops $\gets$ \textbf{op-trans}(op, ReplicaAlgo(ndevs))
     }
     \For{\upshape new\_op, device in \textbf{zip}(new\_ops, env.devices)}
     {
       \textbf{op-assign}(new\_op, device) \\
     }
   }
\caption{Data Parallelism \plcode{}.}
\label{alg:data-parallel-sprogram}
\end{algorithm}

\subsubsection{Interlaced Pipeline Parallelism}
\label{subsubsec:mbart}

% \begin{figure}
%     \centering
%     \includegraphics[width=\linewidth]{figures/moti_newpipeline}
%     \caption{Interlaced Pipeline Schedule}
%     \label{fig:interlaced}
% \end{figure}

\begin{figure}
    \centering
    \includegraphics[width=\linewidth]{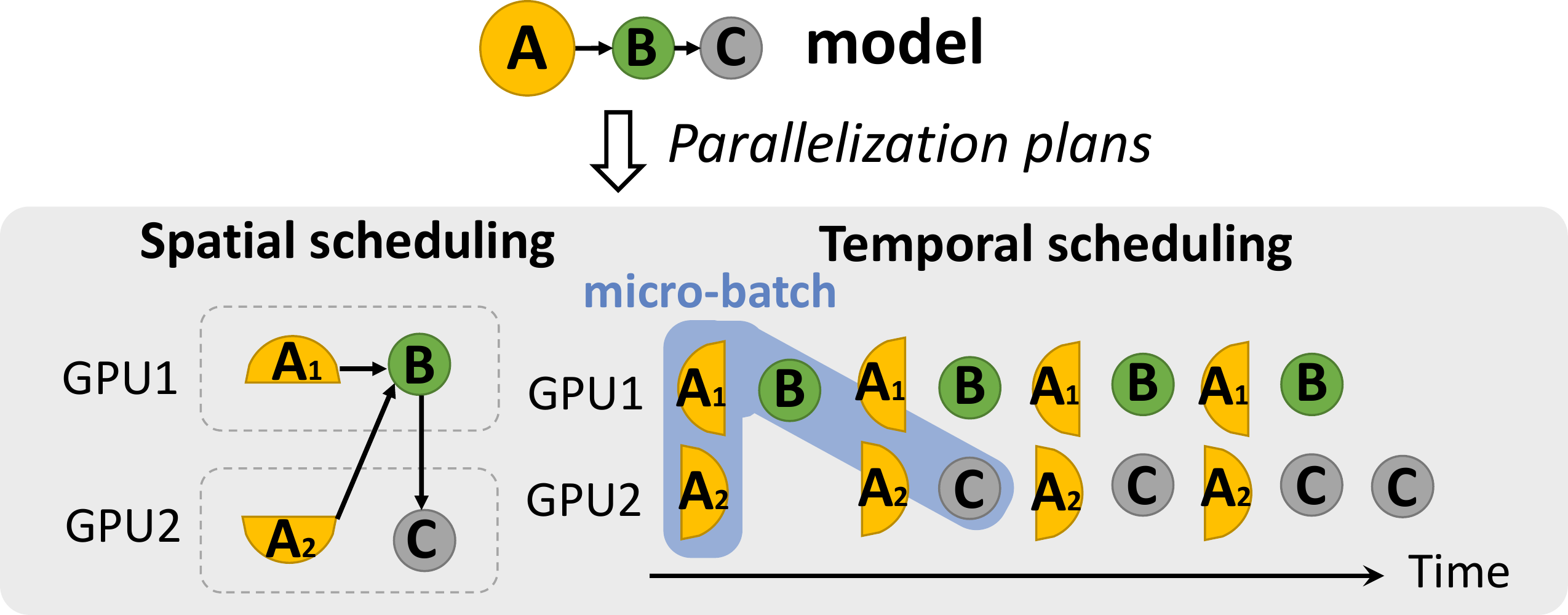}
    % \vskip -0.5ex
    \caption{Interlaced pipeline scheduling.}
    % \vskip -2.5ex
    \label{fig:interlaced}
\end{figure}

mBART is a language translation model~\cite{mbart} with imbalance layers. 
It consists of embedding layers 
and transformer layers.
The embedding layers consume large memory with little computation load, while the transformer layers are opposite,
leading to imbalance resource utilization if organizing the layers into stages.
% Its first embedding layer is connected with multiple encoder layers, followed by another embedding layer connected with a few decoder layers. The embedding layers consume large memory with little computation load, while the encoder and decoder layers are opposite. 
Existing pipeline parallelisms~\cite{megatron1,gpipe} can only place different stages on disjoint devices.
Such parallelization plan will lead to low resource utilization due to imbalanced resource consumption across stages.

To tailor a parallelization plan for this model, we break the assumption of existing pipeline parallelisms that the stages must be placed on disjoint devices. To this end, the embedding layer as the first pipeline stage shares the devices with all other stages. Figure~\ref{fig:interlaced} shows an simplified imbalance model and its tailored parallelization plan. 
Such tailored parallelism can be considered as a regular pipeline among transformer layers, while interlaced with tensor model parallel over embedding layers, which is referred to as \emph{Interlaced Pipeline}.

\para{\plcode{} for Interlaced Pipeline.} Interlaced Pipeline is more complex than existing pipeline parallelisms,
% but with \pn{} the execution plan of this pipeline can be easily expressed as shown in Algo~\ref{alg:interplace}.
%but with flexibility of \pn{}, it can be expressed as 
The corresponding \plcode{} is shown in Algo~\ref{alg:interplace}.
The program first transforms the graph to \texttt{K} micro-batches (Line 2-3). It places transformer operators (\ie \texttt{stage\_ops}) to different devices (Line 6-8). Then, for embedding layers (\ie \texttt{emb\_ops}), it further splits it into \texttt{S} partitions and places them across all devices (Line 10-13).

After operator transformation and placement, the program works on the temporal scheduling (Line 13-22). Transformer operators (\ie \texttt{stage\_ops}) are firstly reordered to follow the same temporal order of 1F1B pipeline~\cite{dapple} into a sequence of stage tasks (Line 13). Then inside a ``for'' loop, \texttt{op-order} is applied to determine such sequential temporal ordering (Line 15-18). Embedding operators (\ie \texttt{embed\_tasks}) are inserted as barriers among transformer operators when the \texttt{step} is a multiple of 2 (Line 19-22).
% It schedules the first micro-batch's embedding layer. Then, for each \texttt{step}, it schedules the next task (\ie \texttt{stage\_tasks}) of the running micro-batches (\ie \texttt{current}). For example, in \texttt{step} 3, \texttt{current} has two micro-batches, the first stage of one micro-batch and the third stage of the other are popped out to be assigned to \texttt{stage\_tasks} (Line 18), they are scheduled after all the previous tasks (Line 19). When the \texttt{step} is a multiply of 2, the corresponding embedding stage (\ie \texttt{embed\_tasks}) is scheduled after the previous tasks (Line 21-25).

\begin{algorithm}[t!]
\DontPrintSemicolon
   \small
   \KwInput{\pn{}Graph g, Environment env, Micro Batch Number $K$}
   % \KwOutput{Transformed Graph $G$}

  $S$ $\gets$ $\lvert$env.devices$\rvert$ \tcp*{number of stages}

  \tcp{==== 1F1B Transformation}

  \For{op $\in$ g.ops}
  {
    dim $\gets$ GetBatchDim(op) \\
    \textbf{op-trans}(op, SplitAlgo(dim, $K$)) \\
  }

  emb\_ops, stage\_ops $\gets$ Classify(g.ops) \\

  \For{sid $\gets$ 1 to $S$}
  {
    ops $\gets$ GetStageOps(stage\_ops, sid) \\
    \textbf{op-assign}(ops, sid)
  }

  \tcp{==== Additional transformation}
  \For{op $\in$ emb\_ops}
  {
    ops $\gets$ \textbf{op-trans}(op, ShardEmbedAlgo($S$)) \\
    \For{device $\in 1$ to $S$}
    {
      \textbf{op-assign}(ops[device], device) \\
    }
  }

  \tcp{==== Interlaced Pipeline Scheduling}

  % micros $\gets$ GatherToMicroBatch(stage\_ops) \\
  tasks $\gets$ OrderTo1F1B(stage\_ops) \\
  % current $\gets$ \{micros[0]\} \\
  previous\_tasks $\gets$ GetEmbedTasks(emb\_ops, 0) \\

  \For{step $\gets$ 1 to 2*(S+K-1)}
  {
    % stage\_tasks $\gets$ PopHeadFromEachMicro(micros) \\
    stage\_tasks $\gets$ PopHeadFromTasks(tasks) \\
    \textbf{op-order}(previous\_tasks, stage\_tasks) \\
    previous\_tasks $\gets$ stage\_tasks \\
    \uIf{step \% 2 = 0}
    {
      embed\_tasks $\gets$ GetEmbedTasks(emb\_ops, step) \\
      \textbf{op-order}(stage\_tasks, embed\_tasks) \\
      previous\_tasks $\gets$ embed\_tasks \\
      % current $\gets$ current $\cup$ \{micros[step/2]\} \\
    }
  }
\caption{Interlaced Pipeline Scheduling}
\label{alg:interplace}
\end{algorithm}

Across all the above examples and other \plcode{} we implemented, we find decoupled primitives of \texttt{op-trans}, \texttt{op-assign} and \texttt{op-order} are expressive enough to cover all parallelization plans (\S\ref{sec:eval}).

\section{Communication Optimization}
\label{sec:comm-opt}

The more flexible parallelization plans may introduce more diverse and unconventional communication patterns. During data dependency materialization, \pn{} optimizes communications in the following ways.

\para{Aligning with efficient communication collectives.} 
Modern communication libraries~\cite{NCCL,gloo,openmpi} usually provide highly efficient, MPI-like collective communication interfaces, \eg broadcast, gather, reduce and all-reduce, which often outperform the peer-to-peer send and receive interfaces.
%Therefore, during data dependency materialization, we 
Hence \pn{} analyzes the data dependency graph and performs a pattern match to replace a group of peer-to-peer communications into high-performance collectives.

\begin{figure}[t]
    \centering
    \includegraphics[width=0.99\linewidth]{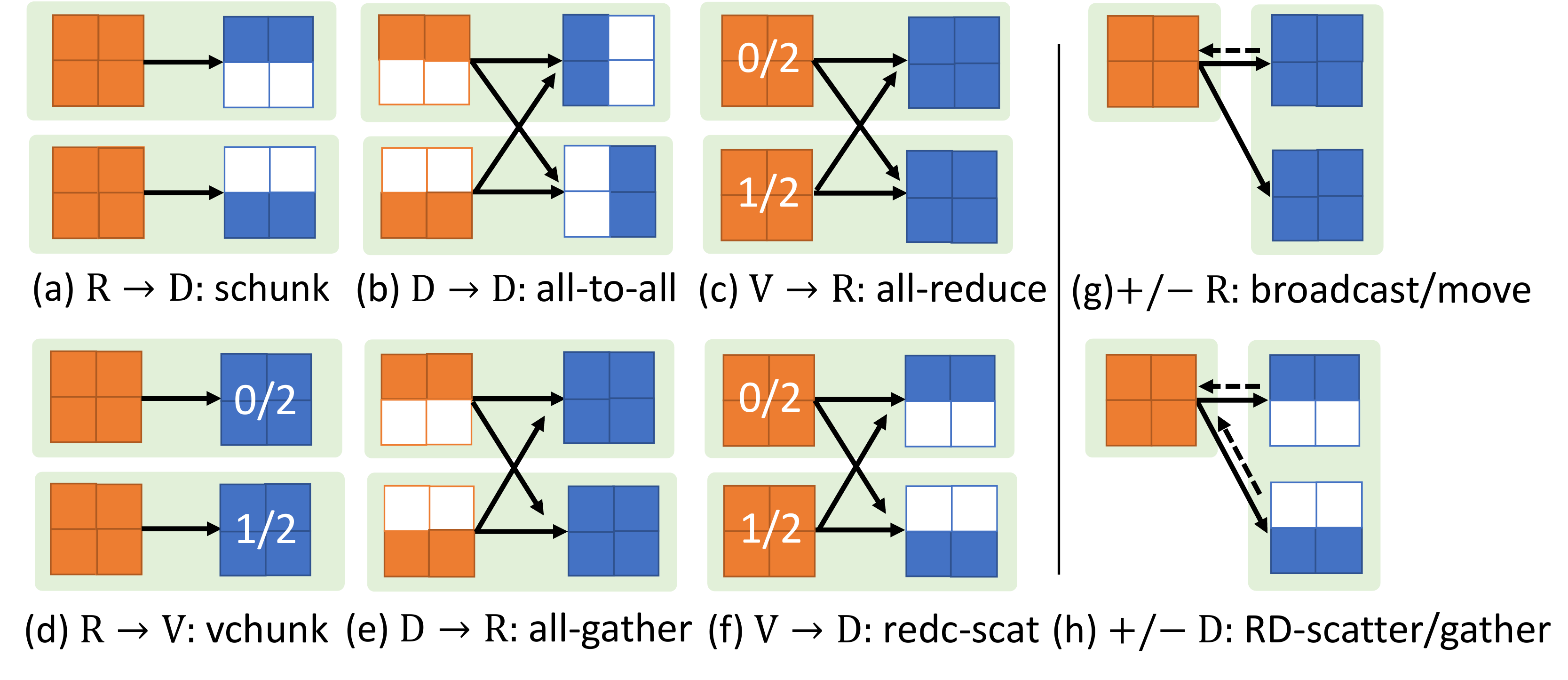}
    \caption{Communication collectives as RVD transitions. A shaded area in (a) to (f) indicates a device. A shaded part in (g) and (h) indicates a device group.}
    % \caption{Collectives used in intra-RVD and inter-RVD, where (g) and (h) are extended collectives for inter-RVD.}
    \label{fig:comm-primitives}
\end{figure}

For complex communication patterns that cannot match any single interface, we design an algorithm to compose the communication with multiple communication primitives based on RVD representation.

\para{RVD representation.} 
DNN clusters are usually equipped with homogeneous accelerator devices~\cite{hived}. Therefore, most parallelization plans partition operators evenly. Thus, their input or output tensors can be simply expressed as: 1)  R($i$), the tensor is replicated to $i$ copies; 2) V($j$), value split, the tensor is decomposed to $j$ copies with the same shape; %(reduce's inverse operation) into $j$ partial value components; 
3) D($k_1$,$k_2$,...,$k_n$), uniformly partition the tensor into $k_1$ parts in the first dimension, $k_2$ parts in the second dimension, so on so forth.
We use RVD to denote the transformation of a tensor.
For example, R(1)V(2)D(1,2) indicates a 2-D pTensor requires no replication, is decomposed into 2 vTensors with the same shape, and each is partitioned into 2 vTensors by partitioning the second axis. Thus, R(1)V(2)D(1,2) can represent 4 vTensors. RVD can represent both producer vTensors and consumer vTensors as they are both transformed from the pTensor.

\begin{figure}[t]
    \centering
    \includegraphics[width=0.85\linewidth]{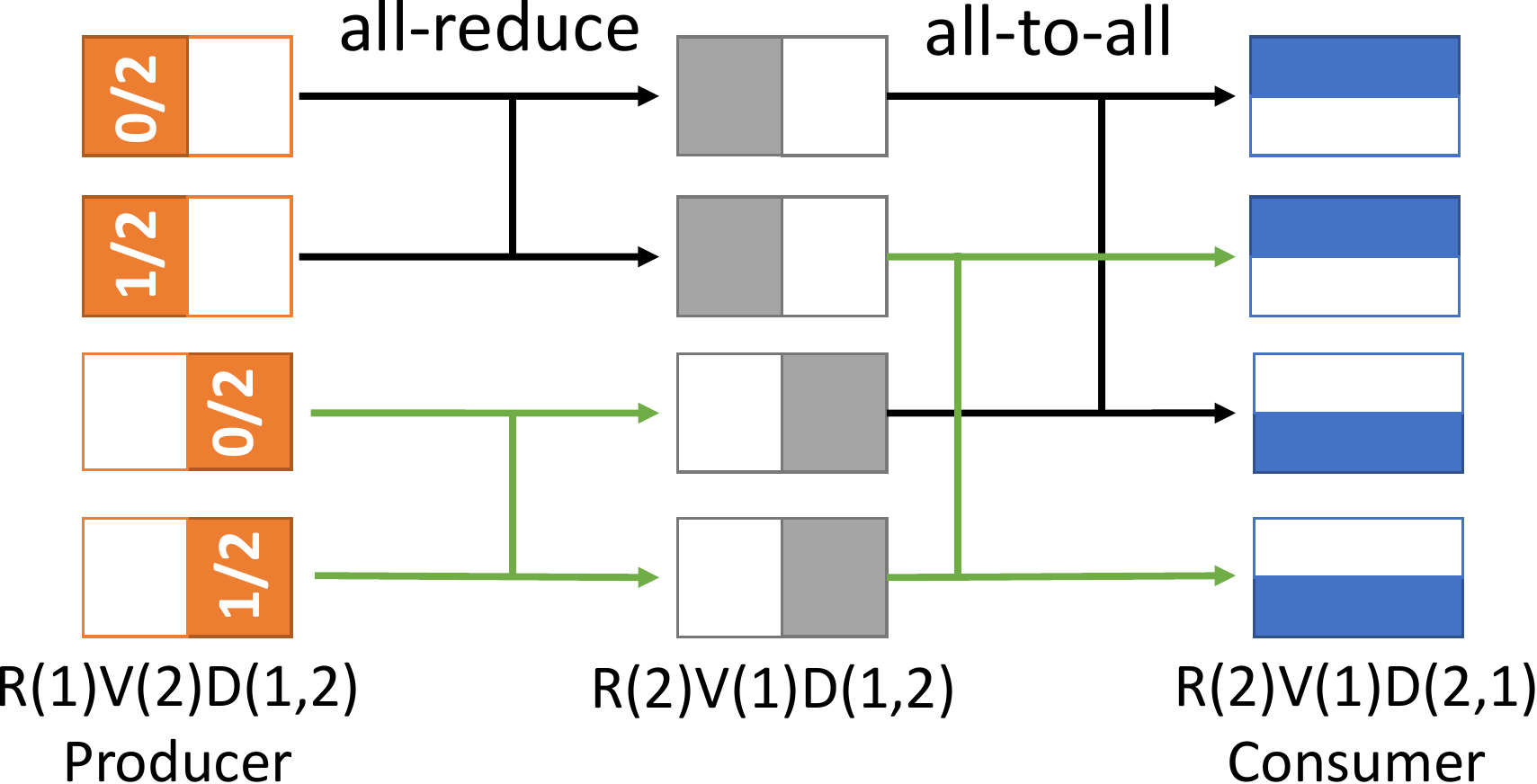}
    \caption{RVD transition path as a dependency materialization plan for an example 2D tensor communication. Arrows with different colors denote different communication groups.}
   \label{fig:rvd-case}
\end{figure}

\para{Communication primitive search over RVD graph.}
Applying a communication primitive essentially turns an RVD to another, with a specific element-wise value exchange pattern (e.g., all-to-all or all-reduce). Thus, a communication primitive defines an RVD transition rule. 
Figure~\ref{fig:comm-primitives}(a - f) lists communication primitives with producers and consumers on the same group of devices, which can be translated into the value changes between R, V, and D ($i / j$ in box indicating value split into $j$ parts and this is $i$-th part).
With the transition rules for different communication primitives as edges, we build an RVD transition graph, and turn the communication composing problem into a problem to find a path from the producer RVD to the consumer RVD.
Figure~\ref{fig:rvd-case} shows an example that connects the producer R(1)V(2)D(1,2) to the consumer R(2)V(1)D(2,1).
% composed communication.
% found by \pn{}. 
It first performs an all-reduce over every 2 tensors to turn V(2) into R(2), and gets R(2)V(1)D(1,2). Then with an all-to-all applied on every two tensors, it converts R(2)V(1)D(1,2) into R(2)V(1)D(2,1).
% remote inter-device RVD
The above solution targets consumer and producer located in the same group of devices, namely intra-device-group RVD, or intra-RVD for short.
It's also possible that producers and consumers locate on different groups of devices, namely inter-RVD. 
For inter-RVD, it first follows the above procedure to build RVD graphs for consumers and producers, respectively. Then it connects the 2 RVD graphs with extended primitives in Figure~\ref{fig:comm-primitives}(g - h) as cross-graph edges, and forms a larger graph.
We assign the edge weight with the time of the communication primitive and leverage Dijkstra~\cite{dijkstra} algorithm to search the shortest path from producer RVD to consumer RVD, and translate the path into a sequence of communication primitives.

This approach can accommodate new communication primitives by formulating a new RVD transition graph.
\section{Implementation}
\label{sec:imple}

We implement \pn{} on top of PyTorch~\cite{PyTorch} with 18k lines of python code.

\para{Converting to \pn{} graph.} We capture the computation graph 
% from one training iteration 
using TorchScript~\cite{torchscript}. Since TorchScript only captures forward graph, we 
complete it with corresponding backward  nodes when converting to \pn{} graph.
% to support more flexible parallelization plans.
% create dummy backward operators if the training iteration contains backward procedure.
% XFG visitor, \eg IsForward(), GetBatchDim()
For convenience, \pn{} also tracks operator types and infers tensor shapes from DFG
% leverages the named tensor~\cite{named-tensor} 
to support functions like \texttt{IsForward()} and \texttt{GetBatchDim()} in \plcode{}.

\para{Autograd for forward operator transformation.} We followed chain rule~\cite{chain-rule} in backward propagation to infer the correct mapping of gradient vTensors. For example, different operators consuming the same vTensor leads to the value-partition of its gradient, which will incur all-reduce for synchronization.
% The gradient of a value-partitioned output needs full-value gradient. 
With this ability, \pn{} can automatically transform backward operators by inferring the gradient vTensors when transforming the corresponding forward operators.

% transformation algorithm
% \para{Providing facilities for primitives.}
% \para{Facilitate transformation.}
\para{Op-trans assistant.}
% There are many operators to represent workload of DL models. Defining transformation algorithms for each operator needs lots of engineering efforts. We leverage the fact that most of operators can follow SPMD partition strategies, that the transformed sub-operators follow the same operator semantic but takes tensors that are replicated or partitioned alongside one dimension.
Defining transformation algorithms for all kinds of operators in DNN models requires lots of engineering effort. Inspired by Einops~\cite{einops}, 
we apply annotations on tensor dimensions to indicate transformation strategies. The annotation is enhanced with keywords to indicate whether a tensor can be partitioned spatially or numerically (value split).

\para{Code generation.} To generate code given a \pn{} graph, the system should specify initial tensor allocation (\eg for model weights). %we consider the initial placement of model parameters outside the scope of graph. 
By default, partitioned weights are initialized to co-locate with its operators. We also allow system experts to specify the initial placement of a part of weights with the tensor slicing interfaces.
% memory optimization
During code generation, we also insert memory-releasing operations, \ie free a python variable, by inspecting tensor life-cycles in the execution plan, which is necessary for inference part to save memory.

\para{Supported parallelization plans.} Table~\ref{tab:support-parallelisms} summarizes our support to existing parallelization rules. %different distributed training techniques from recent work
% parallelization plans from recent work % existing techniques in distributed training
%and checked whether \pn{} can support. 
\pn{} is able to support 15 out of 18 plans. In addition to parallelisms, \pn{} also supports memory optimization techniques such as recompute and swap useful in a parallelization plan. For example, swap can be achieved by an \texttt{op-trans} that inserts an identity operator before and \texttt{op-assign} to CPU. 
% Mechanisms marked with $\ast$ require tailored temporal ordering of communication and computation. 
% For computation workload of these mechanisms, system experts are easy to describe the parallel pattern.
\pn{} by design requires transformation and scheduling to respect the model semantic strictly, thus does not support asynchronous execution in PipeDream.
\pn{} captures graph from one training iteration, thus cannot support cross-iteration scheduling in ByteScheduler~\cite{byte-scheduler}.
\pn{} doesn't access concrete value in tensors, thus cannot figure out data dependency at token level (\ie from a masked tensor) in Terapipe.
% We refer supporting these parallelisms in future work.
We will investigate the support of these parallelisms in future work.

\begin{table}[t]
    \centering
    \small
\begin{tabular}{c|c|c}
    %\hline
    Categories     & Mechanisms    & Support   \\
    \hline \hline
    \multirow{8}*{\shortstack{SPMD\\Parallelism}} 
                                 & Data Parallelism~\cite{data-parallel}         &   $\checkmark\ \ $ \\
                                 & Sequence Parallelism~\cite{megatron3}     &   $\checkmark\ \ $ \\
                                 & Transformer Parallelism~\cite{megatron1} &   $\checkmark\ \ $ \\
                                 & DAP~\cite{fastfold} & $\checkmark\ \ $ \\
                                 & ZeRO~\cite{deepspeed}    &   $\checkmark\ \ $ \\
                                 & Sequence Parallelism~\cite{seq-parallel-colossal}     &    $\checkmark\ast$  \\
                                 % & Multi-D Parallelism~\cite{2d-tp-colossal, 3d-tp-colossal}           &    $\checkmark\ast$  \\
                                 & Flexible Tensor Parallel~\cite{flexflow, gspmd, tofu}  &   $\checkmark\ \ $ \\
    \hline
    \multirow{5}*{\shortstack{MPMD\\Parallelism}}
                                 & 1F1B~\cite{dapple, megatron1} &   $\checkmark\ \ $  \\
                                 & GPipe~\cite{gpipe}            &   $\checkmark\ \ $  \\
                                 & Chimera~\cite{chimera}        &   $\checkmark\ \ $  \\
                                 & PipeDream (Async)~\cite{pipedream}  &   $\times\ \ $   \\
                                 & Terapipe~\cite{terapipe}           &   $\times\ \ $    \\
    \hline
    \multirow{4}*{\shortstack{Memory\\Optimizations}}
                                 & Gradient Accumulation~\cite{ako}  &   $\checkmark\ \ $  \\
                                 & Recompute~\cite{recompute}                &   $\checkmark\ \ $  \\
                                 & Chain-recompute~\cite{DTR}  & $\checkmark\ \ $ \\
                                 & Swap~\cite{huang2020swapadvisor}                     &   $\checkmark\ \ $  \\
    \hline
    \multirow{2}*{Overlapping}
                                 & ByteScheduler~\cite{byte-scheduler}            &   $\times\ \ $  \\
                                 & All-reduce Overlap~\cite{sergeev2018horovod} &   $\checkmark\ast$ \\

\end{tabular}
    % \vskip -1ex
    \caption{The support to existing parallelization plans. `*' denotes additional co-scheduling of computation and communication on system level. Otherwise, the supported parallelisms can be exactly specified by sProgram.}
    % \vskip -2ex
    \label{tab:support-parallelisms}
    \normalsize
\end{table}

% Q0 (\TODO{TBD}): (primitive design) Is it convenient for users to express execution plans?
\begin{comment}
\begin{itemize}
    \item Baseline (Metric: LoC)
    \begin{itemize}
        \item Standalone templates (TP, PP, DP)
        \item Megatron
    \end{itemize}

    \item (LoC) Standalone templates and combination
\end{itemize}
\end{comment}
\section{Evaluation}
\label{sec:eval}

We compare with multiple baseline DNN training systems over diverse real-world models, to demonstrate how \pn{} unleashes performance with new parallelization plans.
Besides end-to-end performance (\S\ref{sec:eval-e2e}), we also analyze the improved memory usage (\S\ref{sec:eval-decouple}), computation efficiency (\S\ref{sec:eval-co-schedule}) and the highly efficient communication generated automatically (\S\ref{sec:eval-comm}).

\subsection{Experimental Setup}

\paragraph{Machine configurations.} 
Our evaluation is performed on a cluster with 32 NVIDIA Tesla V100~(32GB) GPUs. Each server is equipped with 8 GPUs which are connected via NVLink~\cite{NVLINK}. Servers are interconnected with 100~Gbps InfiniBand network.
All the servers are installed with 
% Ubuntu 18.04, NVIDIA driver 450.80, CUDA 11.6~\cite{CUDA}, 
NCCL~2.10~\cite{NCCL} and PyTorch~v1.11.0~\cite{PyTorch}.

\para{DNN models.}
% We evaluate the performance of \pn{} using popular 
We evaluate system performance with four emerging models from diverse domains, including vision, natural language and biology analysis.
1) Swin-Transformer~\cite{swin-v2} is a popular vision model stacked with heterogeneous transformer layers. We set the resolution of input images to 1536$\times$1536, the highest in the paper~\cite{swin-v2}.
2) GPT-3~\cite{gpt-3} is a language model with homogeneous transformer layers. 
% We evaluate GPT-3 for long document tasks
As longer sequences generally perform better~\cite{long-sequence-1, long-sequence-2}, we evaluate GPT-3 for long document tasks with input sequence length of 16384 following LongFormer~\cite{long-sequence-2}'s setting.
3) mBART~\cite{mbart} is a multilingual encoder-decoder model which has a large embedding layer to serve multiple languages in one model. We set input sequence length to 1024 (the default value), and choose the datasets with 500k vocabularies~\cite{vocab500k}.
4) AlphaFold2~\cite{alphafold} is a biological model for predicting protein structures, which has multiple homogeneous evoformer layers.
% AlphaFold adopts a new training pattern that the training data is prepared by performing multiple inference (\ie recycling) in advance. Therefore, each training iteration contains multiple forward and one backward propagation.
We set input with 128 sequences and 256 residues following~\cite{alphafold}, and take a typical training setting: three forward pass
% (\ie the first two are inference) 
and one backward pass in each iteration.

\para{Baseline parallel training systems.} We compare \pn{} with four parallel DNN training systems:
1) Megatron-LM~\cite{megatron2} (v3.0.2) is designed to train transformer-based model, which hierarchically combines pipeline-parallelism with data-/tensor-parallelisms. It evenly partitions model layers into pipeline stages, with all operators in a stage sharing the same data-/tensor-parallelism setting.
2) Alpa~\cite{alpa} (v0.1.5) 
% also hierarchically combines tensor parallelism and pipeline parallelism, and 
further allows 
% fine-grained configurations on each parallelism 
pipeline stages with different data-/tensor-parallelism settings,
and employs a search algorithm to set each stage. 
Among the emerging DNN models we evaluated, GPT-3 is officially supported by Alpa. So, we focus the comparison between \pn{} and Alpa on GPT-3.
For other Alpa officially supported models, \pn{} can follow its search result to build parallelization plans and achieves similar performance. 
% We compare with Alpa on GPT-3 in the evaluation. For other models that are officially supported by Alpa, we skip the comparison due to similar performance.
% which is officially supported among all the evaluated models. We skip the comparison on other models that are officially supported by Alpa due to similar performance.
3) DeepSpeed~\cite{deepspeed} (v0.7.4) applies ZeRO~\cite{zero, zero-offload} optimizations on data parallelism to save memory. %weight memory
% which additionally partition parameters as well as optimizer status across GPU devices and optionally offload them to CPU/NVME memory.
% It doesn't support pipeline parallelism with ZeRO-3 optimization.
% , but allows to combine with manually designed tensor parallelism. 
The highest level memory optimization configuration, ZeRO-3, is not compatible with pipeline parallelism.
% As pipeline parallelism not supported with ZeRO-3 optimization, 
Therefore, we resort to its compatible tensor-parallelism and ZeRO data-parallelism.
% We tune tensor parallelism size and micro-batch size for the best performance. 
As ZeRO-Offload~\cite{zero-offload} memory optimization introduces expensive CPU-GPU communication overhead, we enable it only when memory is insufficient under ZeRO-3.
4) Dynamic Axial Parallelism (DAP)~\cite{fastfold} is a tailored tensor parallelism for AlphaFold2 that partitions the input tensor along a non-batch dimension while replicating weight tensors. 
% It has shown to be more efficient than Megatron-LM. 
We combine DAP with data parallelism (DAP+DP) as baseline for AlphaFold2.
% Since the training pattern in AlphaFold doesn't satisfy the one-forward-one-backward assumption in existing pipeline parallelism, we use DAP with data parallelism as baseline.
%
Hyper-parameters, \eg parallelism size and micro-batch size, significantly affect training performance. 
For a fair comparison, in each evaluation, we tune hyper-parameters for each system to get their optimal settings as baseline.
We also enable recompute~\cite{recompute} 
to save memory only when necessary.
% when memory is not enough.

\subsection{End-to-end Performance}
\label{sec:eval-e2e}

\begin{figure*}[th]
    \centering

    \begin{subfigure}[t]{0.245\textwidth}
        \centering
        \includegraphics[width=0.95\linewidth]{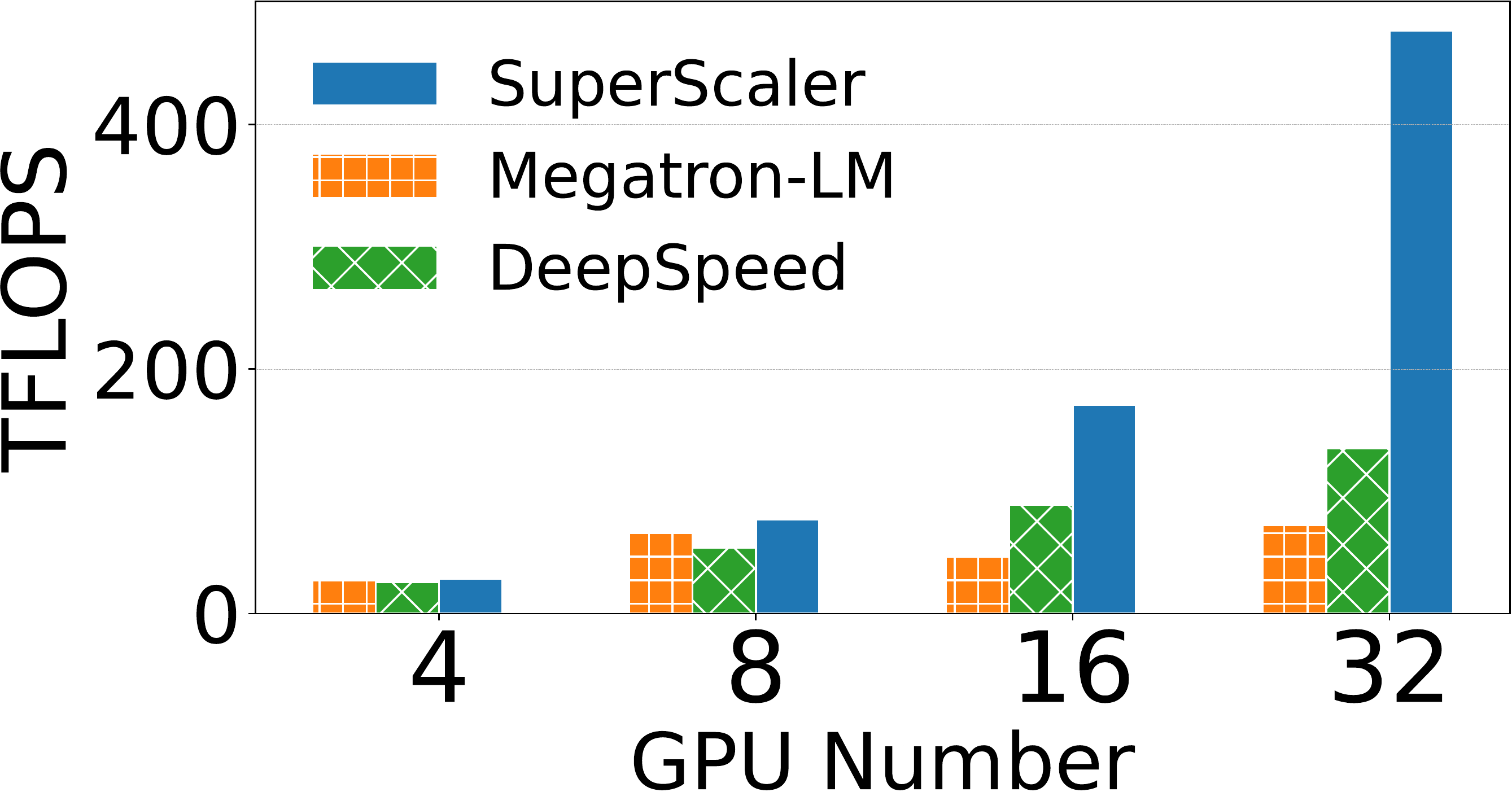}
        \caption{Swin-Transformer}
    \end{subfigure}
    \begin{subfigure}[t]{0.245\textwidth}
        \centering
        \includegraphics[width=0.95\linewidth]{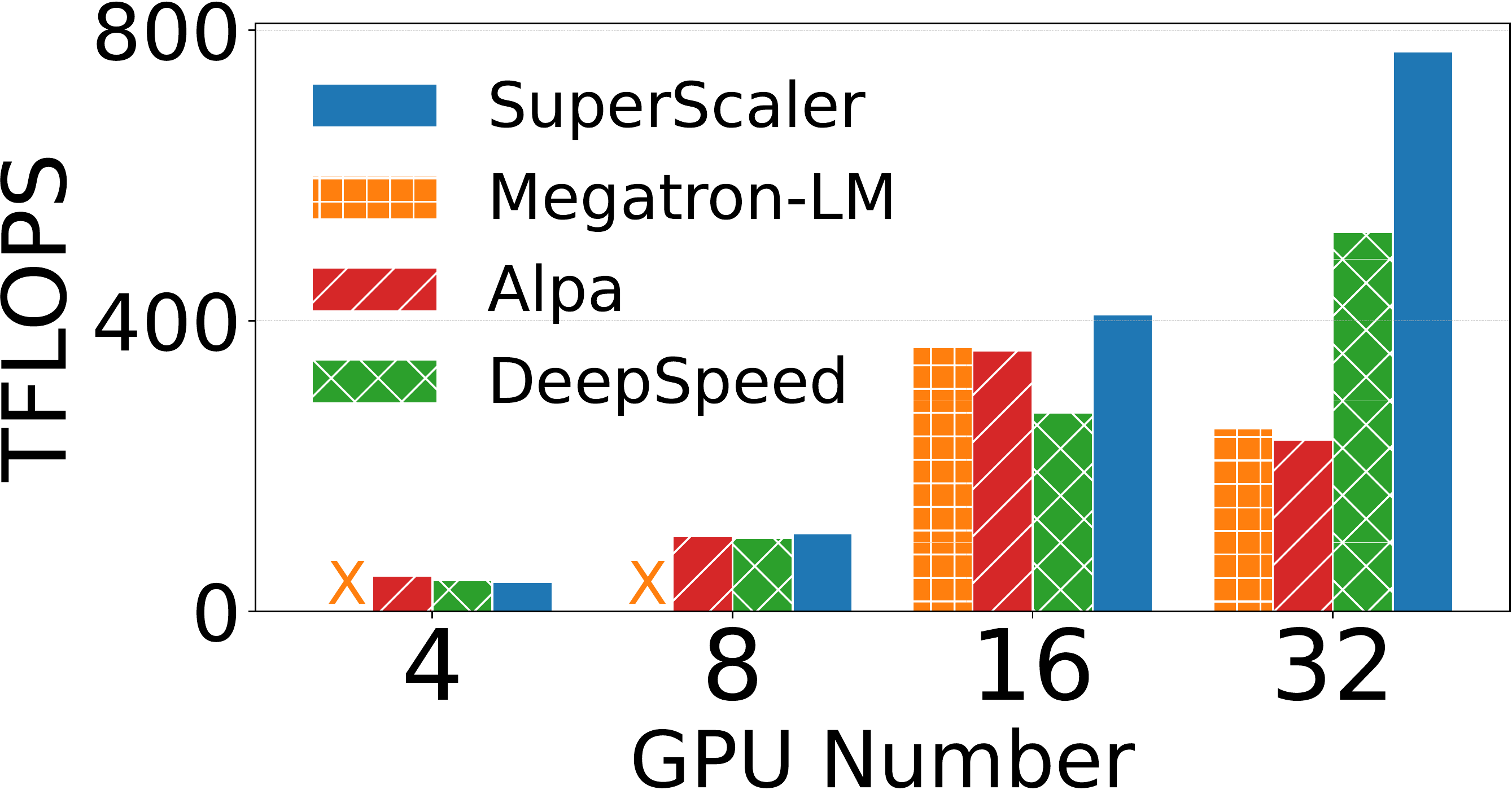}
        \caption{GPT-3}
    \end{subfigure}%s
    \begin{subfigure}[t]{0.245\textwidth}
        \includegraphics[width=0.95\linewidth]{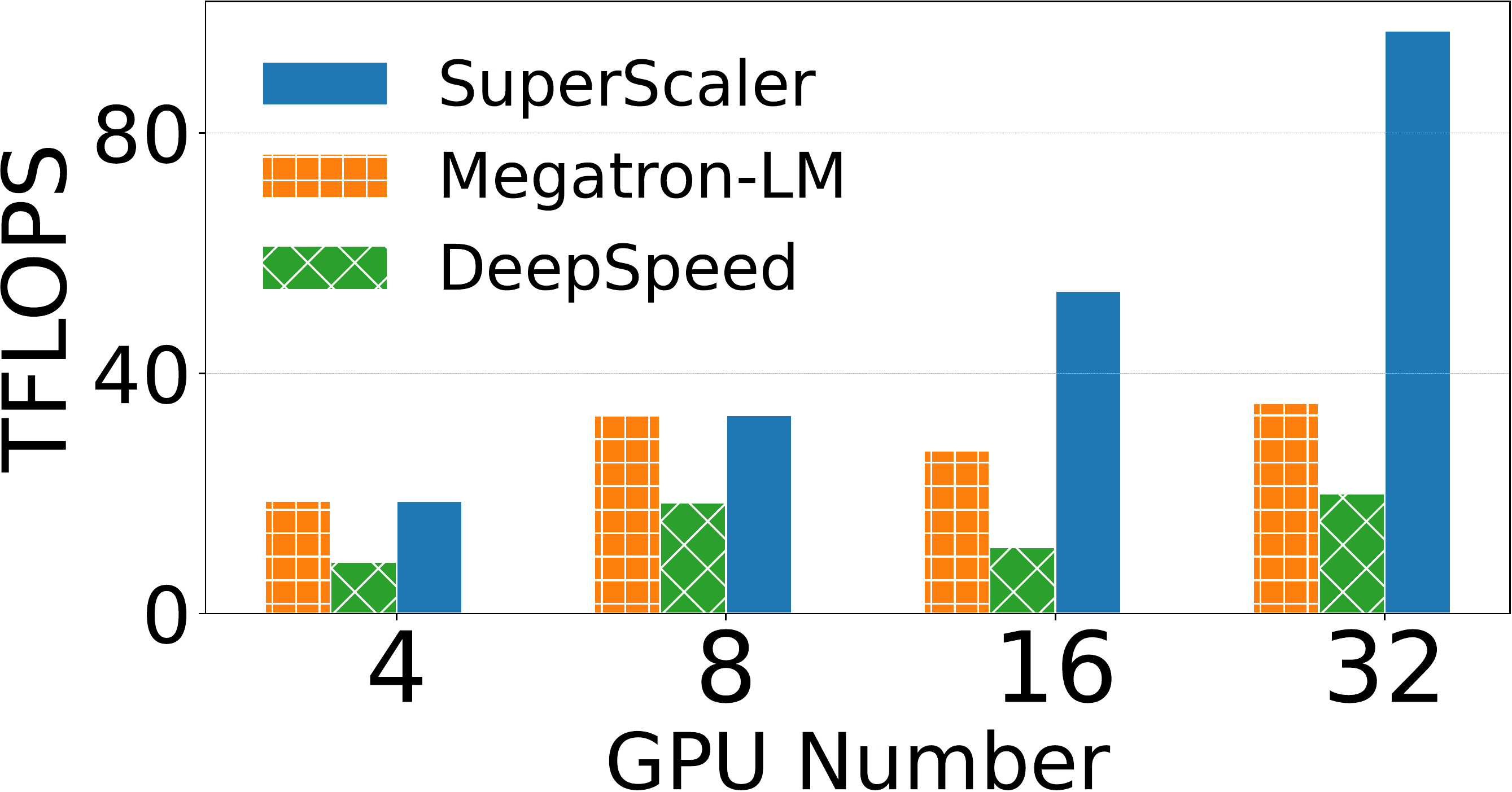}
        \caption{mBART}
    \end{subfigure}
    \begin{subfigure}[t]{0.245\textwidth}
        \includegraphics[width=0.95\linewidth]{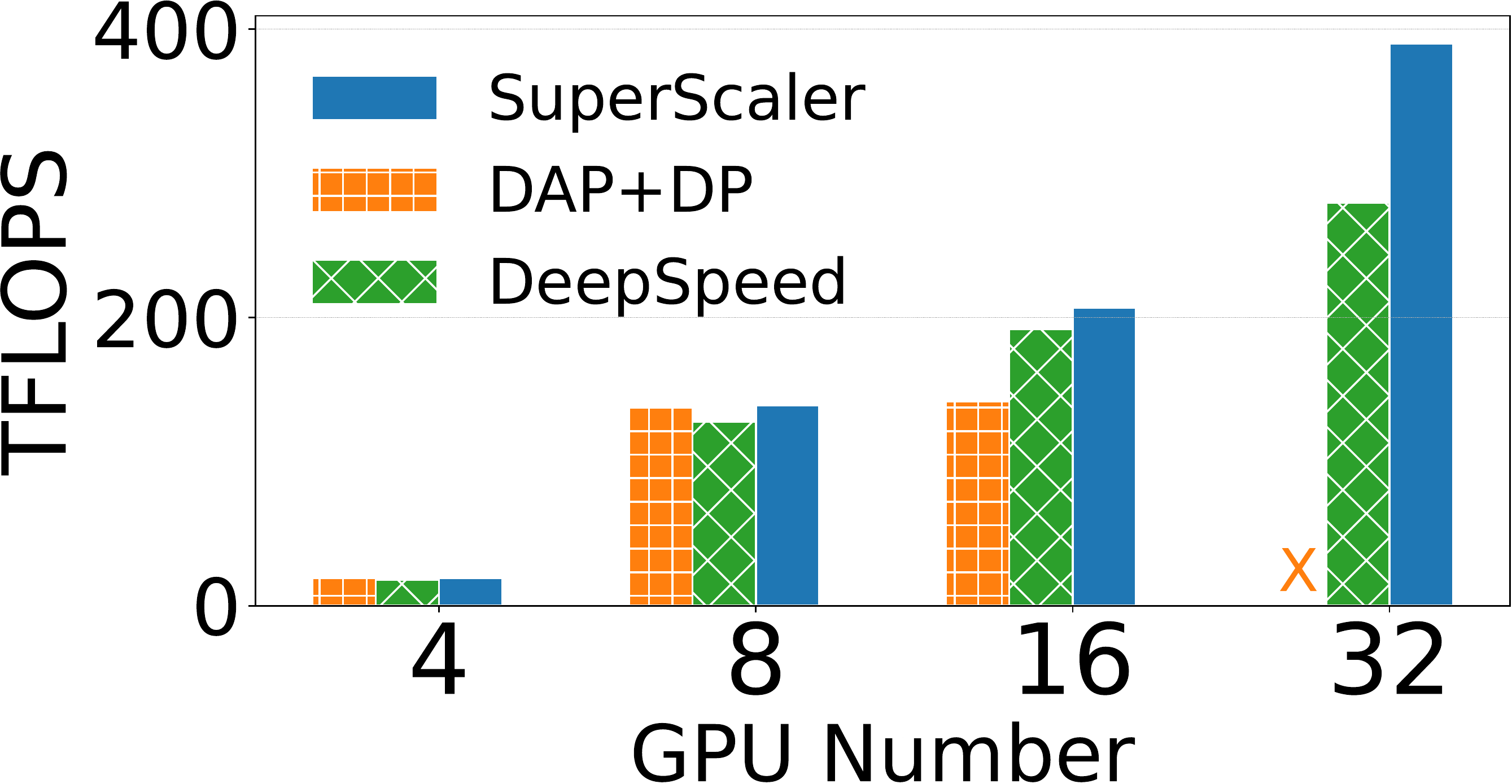}
        \caption{AlphaFold2}
        \label{fig:e2e-alphafold}
    \end{subfigure}
    % \vskip -1ex
    \caption{End-to-end evaluation. $\times$ denotes the failure of training due to out-of-memory.}
    \label{fig:e2e}
\end{figure*}

\begin{table*}[th]
	\centering
	% \footnotesize
	\begin{tabular}{c c c c c c}
		\hline
		\textbf{Model} & \textbf{Parameters} & \textbf{Layer Number} & \textbf{Hidden Size} & \textbf{Head Number}  \\ %& \textbf{GPU Number}\\
		\hline
		Swin-Transformer~\cite{swin-v2} & \{1.8B, 6.6B, 13B, 30B\} & \{32, 48, 56, 64\} & \{512, 768, 1024, 1536\} & \{16, 24, 32, 32\} \\ % & \{1,2,4,8,16,32\} \\
        GPT-3~\cite{gpt-3} & \{1.3B, 2.6B, 6.7B, 15B\} & \{24, 32, 32, 48\} & \{2048, 2560, 4096, 5120\} & \{32, 32, 32, 32\} \\ % & \{1,16\} \\
		mBART~\cite{mbart} & \{4.7B, 9.5B, 20B, 32B\} &
		\{24, 32, 48, 56\} & \{3072, 4096, 5120, 6144\} & \{16, 32, 32, 32\} \\ %& \{2,4,8,16,32\} \\
        AlphaFold2~\cite{alphafold} & \{87M, 930M, 2.4B, 3.2B\} & \{48, 64, 96, 128\} & \{256, 512, 1024, 1024\} & \{8, 16, 32, 32\} \\
		\hline
	\end{tabular}
	% \vskip -1ex
	\caption{Model architecture with increasing number of GPUs. M: million. B: billion.}
	% \vskip -3ex
	\label{tab:model}
\end{table*}

\para{\pn{} parallelization plans.} 
% In addition to the above parallelisms, we combine them with our new discovered techniques to demonstrate the performance improvement. 
\pn{} enables us to apply new parallelization plans over DNN models to improve performance, including \coshard{}~(\S \ref{sec:background}), interlaced pipeline~(\S \ref{subsubsec:mbart}) and 3F1B pipeline schedule~(\S \ref{sec:background}).
For Swin-Transformer and GPT-3, we apply \coshard{} to partition attention heads and feedforward~\cite{attention} hidden dimensions. For GPT-3, it is applied to every transformer layer. For Swin-Transformer, it is only applied to the first four transformer layers which consume the most memory. For mBART, we replace the original 1F1B pipeline with our interlaced pipeline. For AlphaFold2, we use the 3F1B pipeline scheduling.

We perform weak scaling test to evaluate the training performance of all these models. Following the practice of Megatron~\cite{megatron2} and Alpa~\cite{alpa}, we expand model size with the increasingly larger number of GPUs. 
The batch size is set to 128 for AlphaFold2 following~\cite{alphafold}, with other models set to 512.
The complete model configurations are listed in Table~\ref{tab:model}. 
Similar to~\cite{megatron2,alpa}, we evaluate the performance in aggregated tera floating point per second (TFLOPS).
% \TODO{use effective TFLOPS}

\para{Results of Swin Transformer and GPT-3.} Figure~\ref{fig:e2e}(a) and Figure~\ref{fig:e2e}(b) show the end-to-end training performance on Swin Transformer and GPT-3. 
% With \coshard{}, 
Compared to existing baselines,
\pn{} achieves up to 3.5$\times$ speedup on Swin Transformer, and up to 1.5$\times$ speedup on GPT-3.

For Swin-Transformer, \pn{}'s performance is similar to Megatron-LM and DeepSpeed within one server (4-GPU, 8-GPU). This is because the high bandwidth NVLink within a server makes communication difference negligible on both systems, marginalizing \pn{}'s communication improvement.
% have little communication cost comparing with its computation cost. 
DeepSpeed also has a similar performance due to the small model size compared to its heavy computation cost, which can easily overlap with the weight synchronization communication.
On multiple servers (16-GPU, 32-GPU),
\pn{} achieves up to 6.6$\times$ and 3.5$\times$ over Megatron-LM and DeepSpeed, respectively. 
That is because
Megatron's tensor-parallelism
requires more GPUs to fit the larger model size in memory, leading to more communication costs.
%Due to limited inter-server network bandwidth, communication cost is not negligible.
With \coshard{}, \pn{} achieves similar memory usage with less GPUs, showing better performance with reduced tensor-parallelism communication cost.
For example, Megatron-LM requires at least 16-way tensor parallelism in the 16-GPU Swin Transformer test,
while \pn{} only needs a 4-way tensor-parallelism to fit in model, and uses communication efficient 4-way pipeline-parallelism to connect these 4-GPU tensor-parallelism groups.
Similarly, in the 32-GPU test, Megatron-LM requires 32-way tensor parallelism while \pn{} only needs 8-way.

With the reduced memory usage, DeepSpeed also requires less GPUs and outperforms Megatron-LM for similar reason.
% which is also due to its memory saving that allows within-server tensor parallelism. 
However, compared to \pn{}, DeepSpeed still suffers from the extra communication cost due to weights and gradients synchronization.

For the similar reason on Swin-Transformer, on GPT-3, \pn{} achieves up to 3.3$\times$ speedup over Alpa and up to 1.5$\times$ speedup over DeepSpeed. 
% The reason is similar to Swin-Transformer.
For example, we find both Megatron-LM and Alpa need at least 16-way tensor parallelism for 15B model, while \pn{} and DeepSpeed are able to train with 8-way tensor parallelism.
Megatron-LM and Alpa achieve similar performance, which also aligns to the results in their work~\cite{alpa}.
% DeepSpeed outperforms 

\para{mBART results.} In Figure~\ref{fig:e2e}(c), %%CL shows the performance of mBART.
% With interplaced pipeline scheduling,
\pn{} achieves up to 2.8$\times$ and 4.9$\times$ performance speedup over Megatron-LM and DeepSpeed, respectively.
For 4-GPU and 8-GPU cases, \pn{} has the same performance as Megatron-LM, which is because both of them end up with the same parallelization plans,
\ie pure tensor parallelism due to memory constraints. %, thus have same performance.
% For 4-GPU and 8-GPU cases, 
% we find \pn{} has the best performance by setting tensor parallelism size of encoder and decoder blocks to the same number of GPUs, which ends up to the similar parallelization plans found by Megatron-LM, \ie pure tensor parallelism due to memory constraints.
% \pn{} ends up with the same parallelization plans with Megatron-LM, \ie pure tensor parallelism due to memory constraints, thus have same performance.
When model size grows, the embedding layers' weights can no longer fit within a server given tensor parallelism.
Megatron-LM shows worse performance in such cases. It requires the transformer layers co-located with embedding layers must share the same tensor parallelism configuration, which introduces significant communication overheads in transformer layers.
% The template-based solution in Megatron-LM requires encoder and decoder layers that are in the same stage of the embedding layer uses same ways of Tensor Parallelism, thus introduces heavy communications that dominate the training.
Instead, \pn{} only partitions embedding layers across servers, while allows other layers to use within-server tensor parallelism, thus saves expensive inter-server communication.
% In this way, \pn{} saves expensive communication and achieves up to 2.8$\times$ performance speedup over Megatron.
We find DeepSpeed consistently performs less efficiently than Megatron-LM or \pn{}. 
That's because DeepSpeed enables offload, which requires expensive communication to fetch large weights from CPU and synchronize large embedding gradients. 
And such communication can hardly overlap with computation.

\para{AlphaFold2 results.} In Figure~\ref{fig:e2e}(d),  %%CL shows the performance of AlphaFold2. 
\pn{} achieves up to 1.4$\times$ (32-GPU) speedup over baselines with the new 3F1B pipeline scheduling. When model size is small, all systems use data parallelism and therefore performs similarly. With the larger model size,
% when model size gets large, 
both model weights and activations (\ie tensors that between two consecutive forward operators) consume more memory. DAP+DP has to use DAP to partition activations across devices to save memory, which introduces high communication costs and fails to train 3.2B model even with 32-way DAP. Instead, \pn{} leverages 3F1B to partition model weights across devices, which only requires little communication at the boundary of pipeline stages. For example, DAP+DP requires at least 4-way DAP to train the model for 2.4B model, while \pn{} is able to switch from DAP to 4-way 3F1B pipeline scheduling and outperform DAP+DP with 1.5$\times$ speedup on 2.4B model. DeepSpeed can also save memory on weights, but it introduces more overheads by synchronizing weights and gradients than 3F1B scheduling when GPU number is large.
% Therefore, \pn{} outperforms DeepSpeed and achieves up to 1.4$\times$ on 32 GPUs.

\subsection{Reducing Memory with Decoupling}
\label{sec:eval-decouple}

\begin{figure}[]
    \centering
    \includegraphics[width=0.9\linewidth]{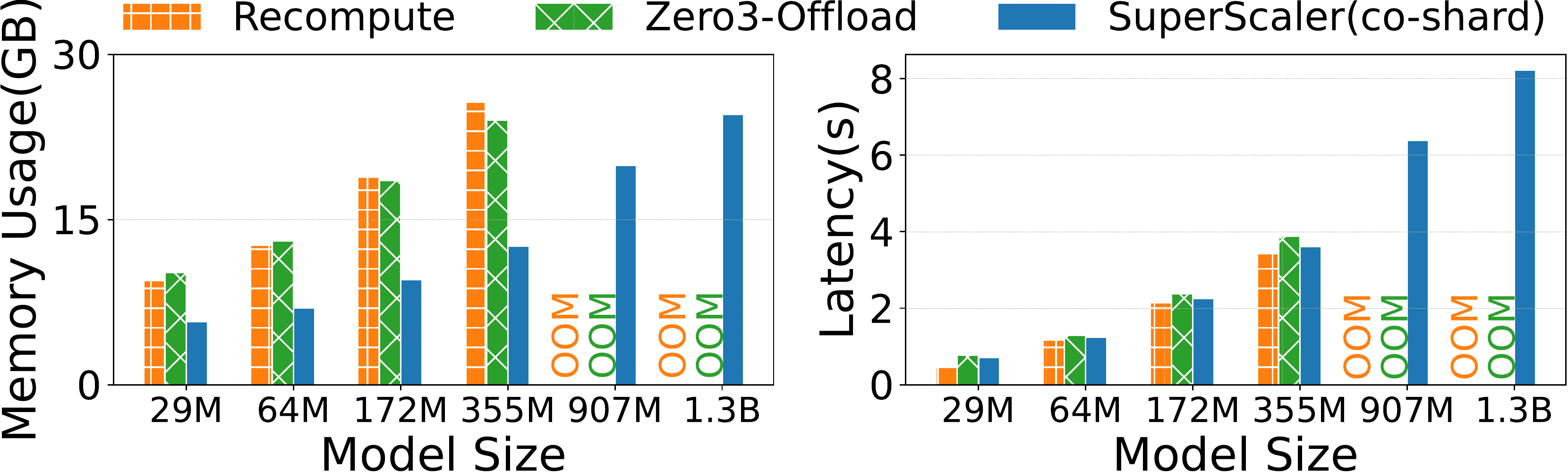}
    % \includegraphics[width=0.95\linewidth]{figures/mo-crop.pdf}
    % \vskip -0.5ex
    \caption{Swin Transformer single-GPU memory consumption and latency with growing model size.}
    % \vskip -1ex
    \label{fig:swin-single}
\end{figure}

\begin{figure}[]
    \centering
    \includegraphics[width=0.9\linewidth]{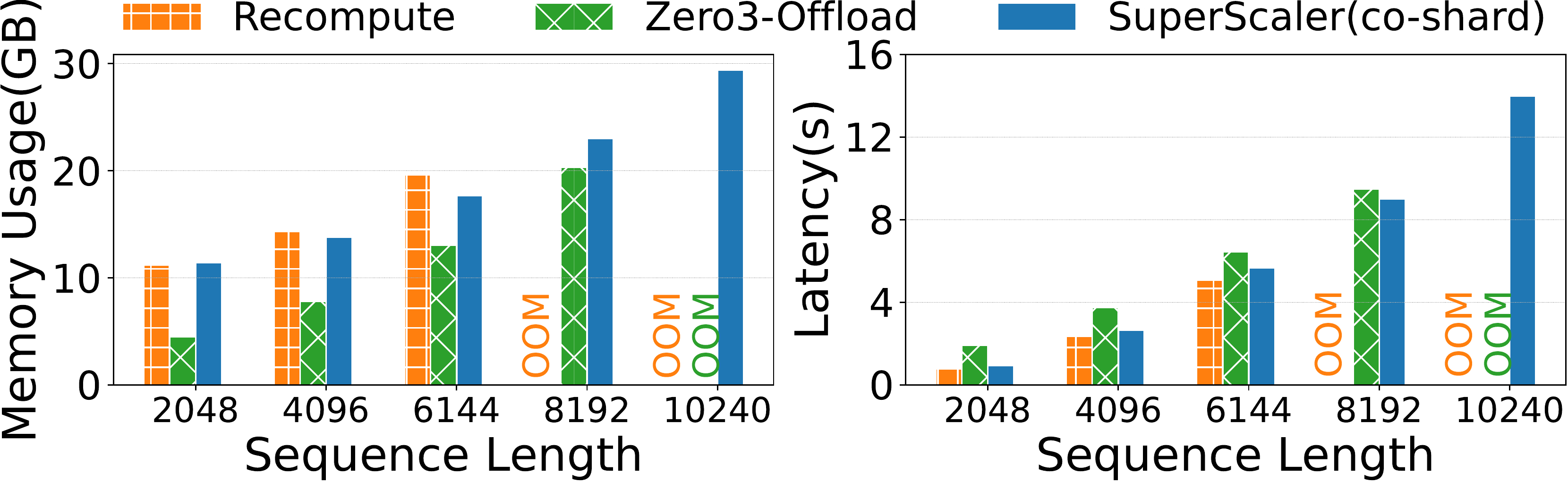}
    \caption{GPT-3 single-GPU memory consumption and latency with growing input size (sequence length).}
    % \vskip -2ex
    \label{fig:gpt-single}
\end{figure}

The decoupling of transformation and scheduling allows parallelization plans to have more flexible scheduling choices of a same transformation, such as reducing memory usage to support training with larger models or larger data samples. To demonstrate this, we take \coshard{} as an example to compare with recompute and ZeRO3-Offload. Both recompute and ZeRO3-Offload are memory-saving techniques. All mechanisms apply recompute for each layer. For \coshard{}, we additionally apply it to attention and feedforward operators. For ZeRO3-Offload, we additionally offload weight and optimizer status to CPU memory. To make fair comparisons on memory consumption, we follow the setting of global batch size 512 but fix the micro batch size to 1.

Figure~\ref{fig:swin-single} shows the single-GPU memory consumption and latency with the increasing model size of Swin-Transformer. With \coshard{}, \pn{} can support training on single GPU with 1.3B model parameters, 3.7$\times$ the model size of ZeRO3-Offload and recompute. Compared to recompute, ZeRO3-Offload can slightly save memory on larger models, but still lead to out-of-memory (OOM.) when model size becomes large (i.e., 907M and 1.3B).
% ends up to the same model size. 
This is because 
ZeRO-Offload can only save
weights and optimizer status memory,
while Swin-Transformer has much larger memory consumption on activations than on weight.
% Swin-Transformer has more larger memory consumption on activations than on parameter, which makes ZeRO-Offload can only save limited memory by offloading parameters and optimizer status. 
In contrast, \coshard{} effectively reduces activation memory by partitioning operators, which reduces the peak memory and hence can train larger models. 

Figure~\ref{fig:gpt-single} shows the single-GPU memory consumption and latency with the increasing input data size, \ie sequence length, of the 1.3B GPT-3 model. \coshard{} supports 1.2$\times$ and 1.7$\times$ the sequence length of ZeRO3-Offload and Recompute, respectively. Compared to Swin-Transformer, GPT-3 usually has relatively larger memory consumption on weights, which explains why ZeRO3-Offload saves more memory than \coshard{} when data size is small. However, ZeRO3-Offload's total memory consumption grows more quickly than \coshard{} with the increase of sequence length, and reaches OOM at the sequence length of 10240. In contrast, \coshard{} has less memory consumption growth rate, and can train with the sequence length of 10240. 

Considering the cost of saving memory by \coshard{} and ZeRO3-Offload, we find \coshard{} breaks large operators into smaller ones, which may lead to less execution efficiency on GPU. However, such impact only slightly slows down the latency, as the model size or input data is large enough to saturate computation. For ZeRO3-Offload, it still suffers from communication overheads when more GPUs are involved, as we observed in \S \ref{sec:eval-e2e}.

\subsection{Improving Efficiency with Co-Scheduling}
\label{sec:eval-co-schedule}

\begin{figure}[t]
    \centering
    \includegraphics[width=0.99\linewidth]{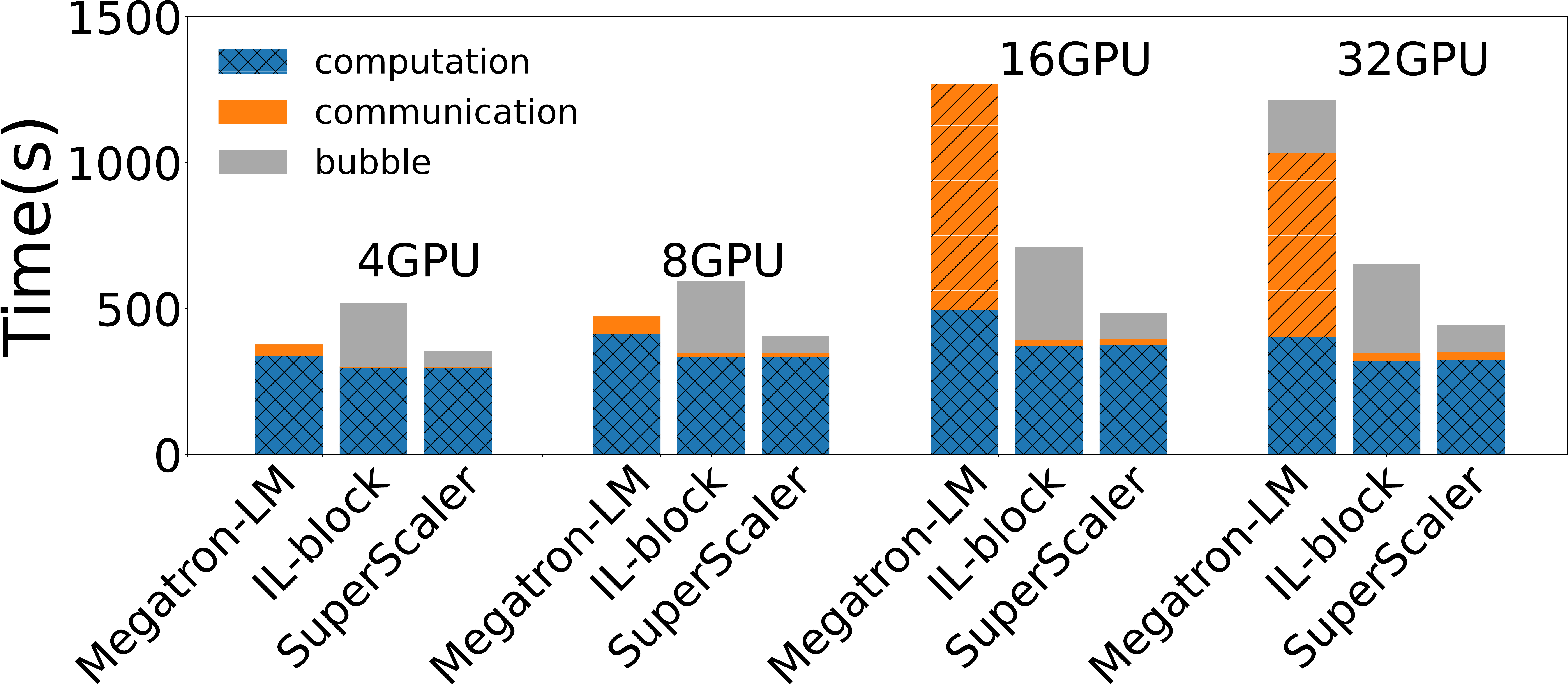}
    \caption{mBART end-to-end performance breakdown.}
    % \vskip -1ex
    \label{fig:mbart-breakdown}
\end{figure}

\begin{figure}[]
    \centering
    \includegraphics[width=0.99\linewidth]{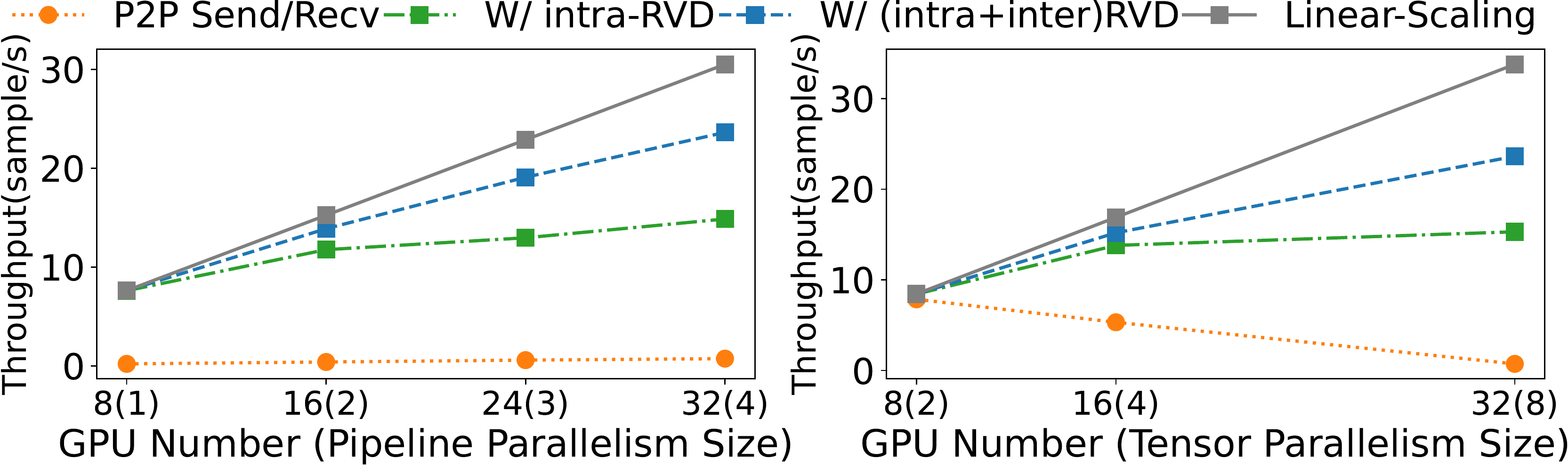}
    \caption{GPT-3 performance with growing pipeline parallelism size (left) and growing tensor parallelism size (right).}
    \label{fig:e2e-comm-e2e}
\end{figure}

The flexible space-time co-scheduling allows \pn{} to explore new parallelization plans for better resource utilization, \eg interlaced pipeline.
We perform a breakdown analysis on mBART to demonstrate how interlaced pipeline improves resource utilization with less idle time and more balanced resource consumption.
% less communication?

We profile the end-to-end training and compare \pn{} with Megatron-LM and Interlaced-block (IL-block). IL-block stands for the interlaced-pipeline scheduling but with the conventional more coarse-grained recompute scheduling. The end-to-end time consists of averaged computation time, communication time and bubble time. Bubble time refers to the device idle time, \eg a stage waiting for data arrival in pipeline parallelism. Figure~\ref{fig:mbart-breakdown} shows the time distribution. 

IL-block achieves up to 1.9$\times$ performance speedup over Megatron-LM. The performance gain mainly comes from reduced communication. 
For 16-GPU and 32-GPU cases, Megatron-LM spends 60\% and 50\% of the total time on expensive cross-server tensor parallelism communication. 
% requires heavy communication cost due to cross-server tensor parallelism, which accounts to 60\% and 50\% time in the whole training time for 16-GPU and 32-GPU cases, respectively. 
In contrast, interlaced-pipeline scheduling only applies cross-server tensor parallelism on embedding layers instead of all layers, which significantly reduces communication cost. % The computation cost is also slightly reduced, which is due to higher computation efficiency of less tensor parallelism in Interlaced scheduling.

Compared to Interlaced-block, \pn{} can achieve another 1.5$\times$ speedup by reducing the bubble time. This is because in Interlaced-block, conventional recompute scheduling treats forward and its backward as a large operator in scheduling, which introduces an unnecessary dependency between the previous backward operation and the forward operation. This leads to the bubble time in the forward operation, waiting for the arrival of gradients. 
In contrast, \pn{} by design follows the fine-grained data dependencies without such unnecessary dependency, which automatically makes the recomputing of forward operation to execute concurrently with previous backward operations. Hence it reduces the GPU idle time and leads to better efficiency. 

\subsection{Serving Diverse Communication Patterns}
\label{sec:eval-comm}

\begin{table}[]
	\centering
	% \footnotesize
        \small
	\begin{tabular}{c c c c}
		\hline
		\textbf{Producers} & \textbf{Consumers} & \textbf{Config. (i$\rightarrow$j)} \\
		\hline \hline
		\textbf{R(i)}V(1)D(1) & \textbf{R(j)}V(1)D(1) & 8$\rightarrow$8, 8$\rightarrow$4, 4$\rightarrow$8 \\
        \textbf{R(i)}V(1)D(1) & R(1)V(1)\textbf{D(j)} & 8$\rightarrow$8, 8$\rightarrow$4, 4$\rightarrow$8 \\
        R(1)\textbf{V(i)}D(1) & \textbf{R(j)}V(1)D(1) & 8$\rightarrow$8, 8$\rightarrow$4, 4$\rightarrow$8 \\
        R(1)\textbf{V(i)}D(1) & R(1)V(1)\textbf{D(j)} & 8$\rightarrow$8, 8$\rightarrow$4, 4$\rightarrow$8 \\
        R(1)V(1)\textbf{D(i)} & \textbf{R(j)}V(1)D(1) & 8$\rightarrow$8, 8$\rightarrow$4, 4$\rightarrow$8 \\
        R(1)V(1)\textbf{D(i)} & R(1)V(1)\textbf{D(j)} & 8$\rightarrow$8, 8$\rightarrow$4, 4$\rightarrow$8 \\
        \hline
	\end{tabular}
	% \vskip -1ex
	\caption{The 18 cases of 6 categories for RVD search benchmark (1-D tensor).}
 % `?' is same with number of producers or consumers, and indicates that the tensor is transformed by replicating or partitioning on that dimension.
	  % \vskip -2ex
	\label{tab:micro-inter-rvd}
\end{table}

\begin{figure}[t]
    \centering
    \includegraphics[width=0.99\linewidth]{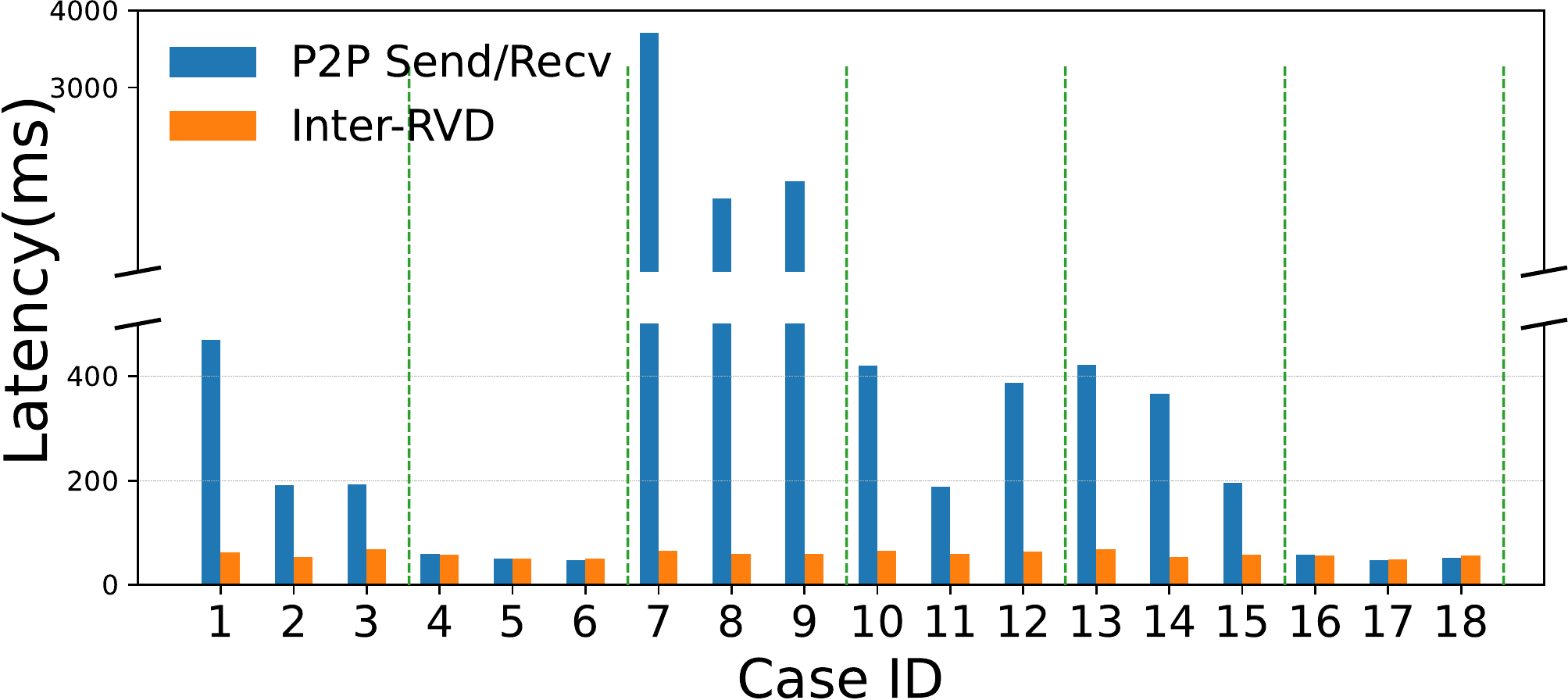}
    % \vskip -1ex
    \caption{Inter-RVD latency on different cases.}
    \label{fig:micro-inter-rvd}
\end{figure}

The flexibility in scheduling may introduce more diverse communication patterns. To evaluate the efficiency of generated communication plans with the RVD representation, we first show the end-to-end training performance of GPT-3 model optimized by intra-RVD and inter-RVD. Then we use micro benchmarks to analyze communication latency of various operator transformation strategies.

\para{End-to-end GPT-3 training.} We use strong scaling to test performance on the 1.3B GPT-3 model.
% We use tensor parallelism and pipeline parallelism to parallelize execution.
We start from using general P2P send/recv (\S\ref{subsec:data-material}) as baseline,
then apply with intra-RVD and inter-RVD step by step.
% which is designed for arbitrary tensor placement without tensor fusion. Based on this, we apply optimizations for intra-RVD and inter-RVD step by step.
% We consider two scaling scenarios: 1) Increasing pipeline parallelism size while freezing tensor parallelism size, where the communication volume of an output tensor is fixed; 2) Increasing tensor parallelism size while freezing pipeline parallelism size, where the communication volume is linearly increased in both intra-RVD and inter-RVD. Figure~\ref{fig:e2e-comm-e2e} shows these two  scenarios.
% Both intra-RVD and inter-RVD benefit more by more GPUs involved in distributed training.
Figure~\ref{fig:e2e-comm-e2e} shows the scalability of training throughput with the growing number of GPUs using: (left) pipeline parallelism (\ie communication message size fixed) and (right) growing tensor parallelism size (\ie communication message size increasing), respectively.
Intra-RVD significantly outperforms P2P send/recv with up to 32$\times$ speedup when involving more GPUs (i.e., 32). Inter-RVD further improves scalability with up to 1.6$\times$ speedup, which is due to reduced cross-server communication costs.

\para{RVD search micro-benchmark.} To evaluate communication search on an RVD graph, we build a micro-benchmark.
As intra-RVD can be viewed as a subset of inter-RVD, we perform inter-RVD search by allocating different numbers of producers on a server, and different numbers of consumers on another server.
% by keeping producers and consumers on a different server with different number. 
Table~\ref{tab:micro-inter-rvd} lists the configurations of producer RVD and consumer RVD. Figure~\ref{fig:micro-inter-rvd} shows the communication latency compared to P2P send/recv solution. Inter-RVD can improve P2P send/recv in 12 out of 18 cases, and achieves up to 57$\times$ speedup. This is because inter-RVD searches communication plans minimize the cost that considers both link bandwidth and communication volume.

\begin{figure}[]
    \centering

    \begin{subfigure}[t]{0.99\columnwidth}
        \centering
        \includegraphics[width=0.99\linewidth]{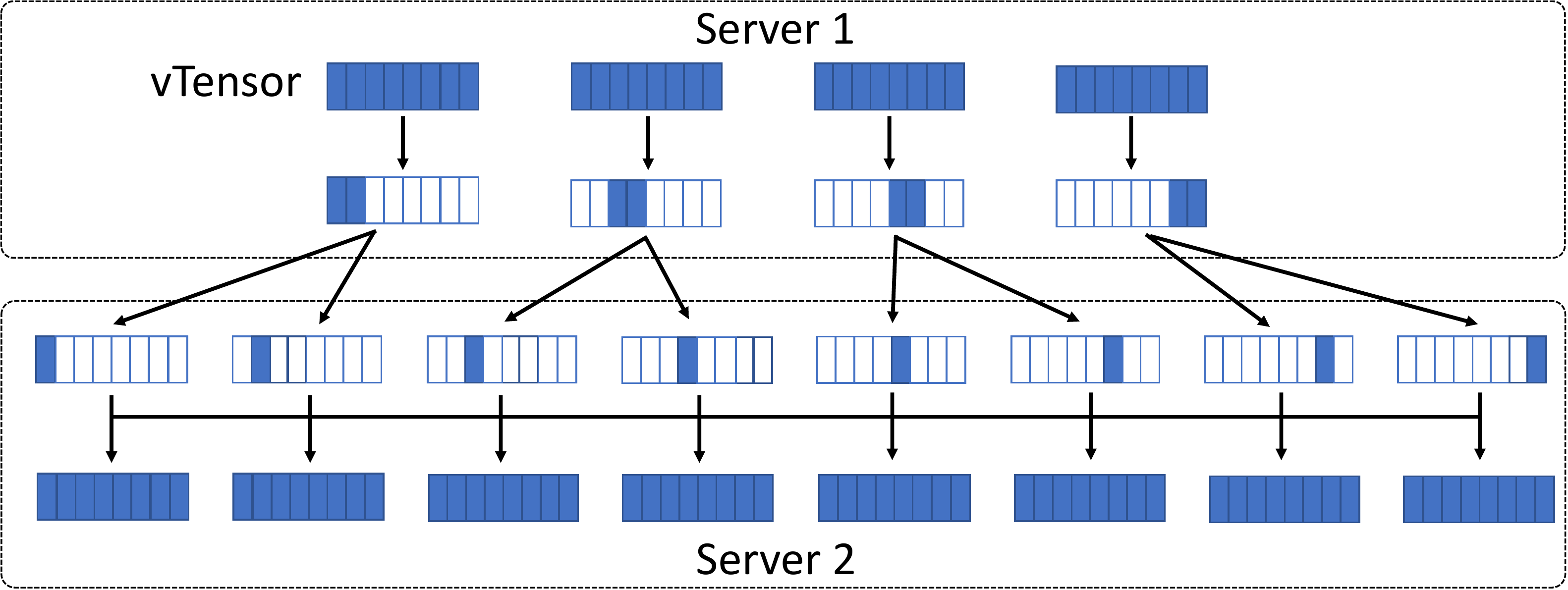}
        \caption{From 4 replicated tensors to 8 replicated tensors.}
    \end{subfigure}
    \begin{subfigure}[t]{0.99\columnwidth}
        \centering
        \includegraphics[width=0.99\linewidth]{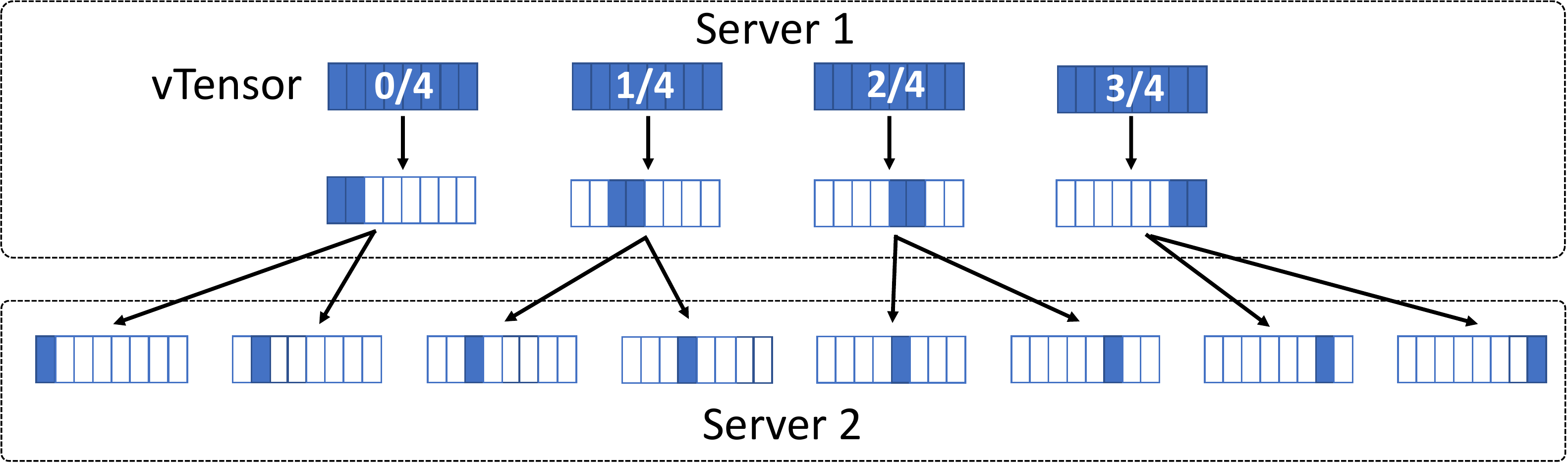}
        \caption{From 4 value-partitioned tensors to 8 axis-partitioned tensors.}
    \end{subfigure}%s
    % \vskip -1ex
    \caption{Two searched cases in inter-RVD.}
    \vskip -2ex
    \label{fig:inter-rvd-case-study}
\end{figure}

\para{Case study.} \pn{} is able to automatically find highly efficient communication plans like manually optimized ones in Alpa and Megatron-LM. 
% We inspect two searched communication plans.
Figure~\ref{fig:inter-rvd-case-study}(a) shows a plan of transferring 4 replicated tensors in Server1 to 8 replicated tensors required by Server2. 
The replicated tensors in Server1 will be partitioned alongside one spatial dimension using \texttt{schunk}~(Figure~\ref{fig:comm-primitives}(a)), and then performs \texttt{RD-scatter}~(Figure~\ref{fig:comm-primitives}(h)) to scatter the sub-tensors to devices on Server2. Finally, Server2 performs an \texttt{all-gather} to gather all tensors. Instead of directly broadcasting tensors inside Server1 to Server2, which is exactly the way of P2P send/recv, RVD search identifies such a pattern due to the low bandwidth across servers, which benefits more by minimizing the communication volume across servers. A similar optimization can be find in~\cite{megatron2}.
Figure~\ref{fig:inter-rvd-case-study}(b) shows another plan of transferring 4 value-partitioned tensors to 8 axis-partitioned tensors. 
Similarly, inter-RVD generates efficient communication plans by first performing \texttt{reduce-scatter} inside Server1, and then using \texttt{RD-scatter} to send sub-tensors to Server2.

\if 0
\subsection{System Overhead}
\label{sec:eval-overhead}

\TODO{only show one or two settings, make conclusion that compile time is related to the number of operators.}

\begin{figure}[t]
    \centering
    \includegraphics[width=0.99\linewidth]{figures/system-overhead-crop.pdf}
    \caption{Time of generating execution plans.}
    \label{fig:system-overhead}
\end{figure}

Figure~\ref{fig:system-overhead} evaluates the total time and its breakdown of generating the execution plan in \pn{} for three models in end-to-end evaluation. \pn{} can generate execution plans within only a few seconds, which is acceptable comparing with large model training time that may last for days or even months. The time of running policy and automatic communication generation increases for larger number of GPUs, which is because that each operator is partitioned into more sub-operators, which bring a larger scope for setting dependency of producers and consumers.

\fi

\section{Related Work}
\label{sec:related}

\lstset{
numbers=none,
basicstyle=\small\ttfamily,
rulecolor=\color{gray},
columns=flexible,
breaklines=true
}

\para{Parallel DNN training.} Recently,
data, tensor, and pipeline parallelisms~\cite{data-parallel, zero, gpipe, dapple, chimera} have been widely used in distributed DNN training.  
In addition, various memory optimizations~\cite{DTR, recompute, huang2020swapadvisor} have been adopted to exploit large-scale model training under GPU memory constraints.
Recently, systems, \eg Megatron-LM~\cite{megatron1,megatron2,megatron3}, DeepSpeed~\cite{deepspeed}, Piper~\cite{piper}, 
Unity~\cite{unity}
and Alpa~\cite{alpa}, 
combine multiple parallelisms and memory optimizations within one system to accelerate distributed DNN training.
However, these solutions fall short in relying on  empirical parallelism configurations and having limited
execution scheduling choices. Thus, despite their successful applications on existing training workloads, they still fail to fully utilize the hardware capabilities. 
In contrast, \pn{} provides a different angle of parallelization, which breaks the building block 
 from predefined parallelisms into fine-grained transformation and scheduling primitives. Thus, \pn{} is able to support more flexible and efficient  parallelization plans, which are considerably crucial for emerging DNN models.

\para{Distributed tensor abstraction and communication.} 
To offer programmability for %%CL extend DNN frameworks for 
parallel execution, distributed tensor abstraction~\cite{pytorch-dtensor, tf-dtensor, oneflow} is introduced to annotate %%CL express 
tensors with partitioning strategies and device mappings, often combined with 
% single-program-multiple-data (SPMD) parallelisms, \eg 
%%CL combination of 
data or tensor parallelism.
However, tensors partitioning annotations only label transformation on operators, 
thus cannot express further flexible scheduling among partitioned parts. \pn{} incorporates transformed vTensor and operators, allowing flexible space-time scheduling, while keeping data dependency tracking.
Furthermore, %%CL with tensor or operator partitioning, 
various communication methods need to be injected between two consecutive operators when they operate on tensors with different partitioning methods. Due to their empirical parallelisms, %%CL According to combined parallelisms patterns, 
systems like Mesh-TensorFlow~\cite{tensorflow-mesh}, tofu~\cite{tofu}, FlexFlow~\cite{flexflow} and Megatron~\cite{megatron2} rely on a few rules to generate such required communication methods. 
GSPMD~\cite{gspmd}, OneFlow~\cite{oneflow} and~\cite{array-redist,pipeline-comm} additionally  investigate communications for different tensor partitioning in pipeline parallelism.
However, they cannot fully satisfy \pn{}'s demand given its numerous types of transformation and arbitrary scheduling.

\para{Parallelization plan search.} To improve the training performance with combined parallelisms, DNN systems~\cite{alpa, piper, tofu, flexflow, pesto, autosync, paddlepaddle} using different searching techniques to find efficient parallelism configurations. 
Most recently, Alpa~\cite{alpa} leverages both integer programming and dynamic programming solvers.
\pn{} is an parallelization plan engine that focus on expanding the parallelization space beyond combined parallelisms, and complementary to above algorithms.

There are some other systems that can work together with \pn{}, including DNN execution systems like Pathways~\cite{pathways} and on device execution compilers~\cite{tvm, rammer,xla}.
%\vskip -2ex
\section{Conclusions}
\label{sec:conclude}

\pn{} is a generation engine of parallelization plans used for parallel deep learning training. We observe that existing training systems rely on empirical parallelization plans, which jointly consider multiple intertwined factors at the same time. This practice shows limited flexibility and leads to the exclusion of promising parallelization plans. To overcome this, \pn{} 
decouples the intertwined factors of model partitioning, scheduling, and data dependency preserving and enables the expression of highly flexible parallelization plans.
As a result, \pn{} can not only support existing popular parallelization plans, but also explore new ones that significantly improve the training performance of emerging models as well as well-optimized large language models.

%-------------------------------------------------------------------------------
%\bibliographystyle{plain}
%\bibliography{bib/papers}

%%%%%%%%%%%%%%%%%%%%%%%%%%%%%%%%%%%%%%%%%%%%%%%%%%%%%%%%%%%%%%%%%%%%%%%%%%%%%%%%
\end{document}